\documentclass[12pt]{iopart}
\usepackage{iopams}

\expandafter\let\csname equation*\endcsname\relax
\expandafter\let\csname endequation*\endcsname\relax
\usepackage{graphicx}
\usepackage{times}
\newcommand{\qqq}{\end{document}}
\DeclareMathAlphabet{\mathcal}{OMS}{cmsy}{m}{n}
\usepackage{xcolor}

\usepackage{bbding}
\usepackage{mathrsfs}
\usepackage{braket}
\usepackage[protrusion=true,expansion=true]{microtype}
\usepackage{cite}
\usepackage{dsfont}
\usepackage{centernot}
\usepackage{color,amsthm,amsmath,amsxtra,amsfonts,dsfont,graphicx,bm}
\usepackage[colorlinks=true,linkcolor=blue, citecolor=blue, urlcolor=blue, bookmarks]{hyperref}
\usepackage{graphicx}
\usepackage{tikz}
\usepackage{braket}
\usepackage{multirow}
\usepackage{comment}
\usepackage{makecell}

\usepackage{booktabs}
\usepackage{tikz}

\newcommand{\dd}{\mathrm{d}}
\newcommand{\ii}{\mathrm{i}}
\newcommand{\EE}{\mathbb{E}}

\newcommand{\OSEE}{S_{\mathrm{op}}}
\newcommand{\cG}{\mathcal G}
\newcommand{\SO}{\mathrm{SO}}

\newcommand{\PV}{\operatorname{PV}}

\begin{document}

\title{Operator-Space Entanglement Entropy in Random Fermionic Gaussian Dynamics}

\author{
Pablo Bayona-Pena$^{a,b,\dagger,*}$,
Elena Gramellini$^{a,\dagger}$,
Lorenzo Piroli$^{a,b}$
}

\address{$^a$Dipartimento di Fisica e Astronomia,
Universit\`a di Bologna,
Via Irnerio 46,
40126 Bologna, Italy\\[4pt]
$^b$INFN, Sezione di Bologna,
Via Irnerio 46,
40126 Bologna, Italy\\[4pt]
$^\dagger$These authors contributed equally to this work.\\[4pt]
$^*$Author to whom any correspondence should be addressed; ~\href{mailto:juanpablo.bayonapen2@unibo.it}{juanpablo.bayonapen2@unibo.it}
}

\begin{abstract}
We study the von Neumann operator-space entanglement entropy (OSEE) of initial product operators evolved under random gaussian dynamics, as modeled by a Brownian SYK$_2$ model. By exploiting a mapping to a domain-wall melting problem, we derive an exact expression for the time-dependent bipartite OSEE in the thermodynamic limit, and obtain an analytic characterization of the corresponding gaussian Page curve emerging at late times. Our results build on recent work on the entanglement dynamics in the all-to-all quantum simple exclusion process, leveraging tools from RMT, including the Coulomb-gas approach and a mapping to the free Jacobi process. Finally, based on our results, we discuss the possibility of relating the OSEE with non-gaussianity measures in the Heisenberg picture.
\end{abstract}

\section{Introduction}
\label{sec:intro}

The operator-space entanglement entropy (OSEE)~\cite{Zanardi2001,ProsenPizorn2007,hartmann2009density,Dubail2017} extends to operators the notion of bipartite entanglement entropy. Formally, one maps the operator onto the corresponding Choi state~\cite{nielsen2000quantum} in a Hilbert space where the number of degrees of freedom is doubled compared to the original space, and defines the OSEE as the entanglement entropy of that state. 

From the point of view of many-body physics, the OSEE is important as it allows us to extend to operators a notion of classical simulability that was first developed for quantum states. For instance, it is by now a standard result that one-dimensional states with low bipartite (R\'enyi) entanglement entropies can be approximated efficiently by matrix-product states (MPS)~\cite{verstraete2006matrix,schuch2008entropy} and thus simulated on a classical computer~\cite{schollwock2011density}. In 1D, an analogous result states that low OSEE (under certain conditions on the operator) implies the existence of an efficient representation in terms of matrix-product operators (MPOs)~\cite{Dubail2017,cirac2017matrix}.

The OSEE is generally more suitable to make statements about the complexity of quantum dynamics compared to the state entanglement entropy, since the former may remain small for the initial operators of interest, while the latter is generically growing linearly in time, away from very special cases~\cite{ProsenPizorn2007,Dubail2017}. In fact, the general expectation is that, under certain hypotheses, the growth of OSEE should distinguish between integrable~\cite{calabrese2016introduction,essler2016quench} and chaotic~\cite{d2016quantum} dynamics~\cite{prosen2007efficiency,muth2011dynamical,hosur2016chaos,ZhouLuitz2017,bertini2020scrambling,styliaris2021information}. For instance, the bipartite entanglement entropy of local observables for local, free-fermionic dynamics is bounded by a function growing at most logarithmically in time~\cite{ProsenPizorn2007}, while it is expected that, for chaotic local Hamiltonians, it grows linearly~\cite{vonKeyserlingk2018operator,nahum2018operator,bertini2020operator,bertini2020operator_II} (see also Refs.~\cite{alba2019operator,alba2021diffusion,pozsgay2026gsquare} for recent work on the OSEE of interacting integrable dynamics).

Despite the importance of the OSEE, several questions remain open. For instance, for non-local observables and dynamics beyond 1D, the behavior of OSEE is generally less clear. In addition, recent work has started to explore the relation between the OSEE and different notions of complexity: notably, Ref.~\cite{DowlingModiWhite2025} put an upper bound on the OSEE in terms  of the quantum resource of non-stabilizerness in operator space~\cite{VeitchEtAl2014,chitambar2019quantum,liu2022many,DowlingKosTurkeshi2025}. While these questions are interesting, a precise answer is often limited by the difficulty of computing or estimating the OSEE beyond integrable dynamics.

These observations motivate extending the study of the OSEE to analytically tractable models in the absence of local conservation laws.  In this work, we consider fermionic systems and initially unentangled Majorana strings, \emph{i.e.} products of local Majorana operators, and study the evolution of the corresponding OSEE under random gaussian evolution, as modeled by a Brownian SYK$_2$ model~\cite{saad2018semiclassical,SunderhaufPiroliQi2019,jian2021note,agarwal2022emergent} (after Sachdev, Ye, and Kitaev~\cite{sachdev1993gapless,maldacena2016remarks}). While the dynamics is gaussian and thus classically efficient to simulate~\cite{Bravyi2005,surace2022fermionic}, the space-time disorder destroys local conservation laws, deviating from previous work. After a mapping to a domain-wall (DW) melting problem, we extend recent tools~\cite{BernardPiroli2021,BernardPiroliScopa2026} developed in the study of the bipartite entanglement in the quantum simple exclusion process~\cite{bauer2017stochastic,bauer2019equilibrium,barraquand2025introduction}, and obtain two main results. First, we provide an analytic expression for the time evolution of the bipartite entanglement entropy at all times in the thermodynamic limit. Second, we compute gaussian Page curves parameterized by both the subsystem fraction and the length of the initial Majorana string. These curves differ qualitatively from the Page curve of fully Haar-random many-body states~\cite{Page1993}. In addition, finite size corrections are also distinct from previously studied Page curves in random fermionic gaussian pure states~\cite{BianchiHacklKieburg2021}.

Our results provide a detailed description of OSEE in random gaussian dynamics, adapting techniques from random-matrix theory (RMT)~\cite{forrester2010log} that could be of independent interest. In addition, they also have implications for the relation between OSEE and another quantum resource, the (fermionic) non-gaussianity~\cite{genoni2008quantifying,genoni2010quantifying,marian2013relative,lumia2024measurement,ares2026asymmetry,turkeshi2026williamson}, in the Heisenberg picture. In particular, we discuss possible constraints that our results put on an upper bound of the OSEE in terms of a natural extension of non-gaussianity to operator space.

The rest of this manuscript is organized as follows. In Sec.~\ref{sec:preliminaries}, we introduce the preliminary tools needed for our analysis, including fermionic gaussian states and the operator-state mapping. We also carefully define the model of random gaussian dynamics studied in this work. Secs.~\ref{sec:page_curves} and~\ref{sec:OSEE_dynamics} are devoted to the analysis of the asymptotic stationary value of the bipartite entanglement entropy and its dynamics, respectively. In both cases, we adapt previously developed RMT tools to obtain fully analytic results. In Sec.~\ref{sec:non_gaussianity}, we discuss implications of our work for the possibility to put an upper bound on the OSEE in terms of a notion of non-gaussianity in the Heisenberg picture. We conclude in Sec.~\ref{sec:conclusions}, which also includes an outlook on future directions, while the most technical derivations are consigned to the appendices.

\section{Preliminaries}
\label{sec:preliminaries}
In this section we introduce the tools to investigate the OSEE of fermionic operators and the model of random gaussian dynamics studied in this work. 

\subsection{Gaussian states and entanglement entropy}
\label{sec:Gaussian states and entanglement entropy}

We begin by introducing gaussian states and the operator-space mapping for fermions. In the following, we will consider a system of $N$ physical fermionic modes, that correspond to $L=2N$ Majorana operators. Denoting by $\{c_j\}_{j=1}^N$ the fermionic annihilation operators, the Majorana modes are defined by
 \begin{equation}
 \psi_{2j-1}= c_j + c^\dagger_j, \quad  \psi_{2j}=- i( c_j - c^\dagger_j).
 \end{equation}

A state $\rho$ is said to be gaussian if it satisfies Wick's theorem~\cite{Bravyi2005}. If $\rho$ is mixed and full-rank, this is equivalent to saying that $\rho$ can be expressed as the exponential of a quadratic Hamiltonian in the modes $\{c_j,c_j^\dagger\}_{j=1}^N$, $\rho\sim e^{-(1/2) c^T Hc}$, where $c=\{c^\dagger_1,\cdots,c^\dagger_N, c_1,\cdots,c_N \}$. Because of Wick's theorem, all correlation functions of gaussian states are completely specified by their two-point correlation functions~\cite{Bravyi2005}, which are encoded in the so-called covariance matrix
\begin{equation}
    \mathcal C_{jk}(|\psi\rangle)=-\frac{i}{2}\langle \psi|\left[ \psi_j, \psi_k\right]|\psi\rangle .
    \label{eq:covariance matrix general}
\end{equation}
The matrix is real and antisymmetric and its eigenvalues are purely imaginary and come in pairs $\{\zeta_j\}_{j=1}^{2N}=\{i\nu_j\}_{j=1}^N\cup \{-i\nu_j\}_{j=1}^N$.

Remarkably, the covariance matrix also uniquely specifies the bipartite entanglement entropy. Given $|\psi\rangle$ and a bipartition of the system into a region $A$ and its complement $B$, we recall that the bipartite (von Neumann) entanglement entropy is defined by
\begin{equation}
    S_{A}= - {\rm Tr}[\rho_A \log \rho_A], \quad \rho_A= {\rm Tr}_B[\rho]\,.
\end{equation}
For a gaussian state associated with the covariance matrix~\eqref{eq:covariance matrix general}, we have~\cite{vidal2003entanglement}
\begin{equation}\label{eq:h_2_formula}
    S_A=\sum_j {\rm h}_2\left(\frac{1+\nu_j}{2}\right),
\end{equation}
with 
\begin{equation}
    \text{h}_2(x)= -x\log x -(1-x)\log(1-x).
\end{equation} 
Here, $\pm i\nu_j$ are the eigenvalues corresponding to the reduced covariance matrix $\mathcal{C}_A$, which is obtained by only selecting $j,k\in A$ in Eq.~\eqref{eq:covariance matrix general}. Note that the sum in Eq.~\eqref{eq:h_2_formula} runs over one member in each pair.

Moreover, for number conserving states we can equivalently consider the complex fermion covariance matrix
\begin{equation}
    G_{jk}(|\psi \rangle) = \langle \psi| c_j^\dagger c_k|\psi\rangle .
    \label{eq:covariance_matrixU1}
\end{equation}
If $\lambda_j$ is an eigenvalue of $G$, the corresponding $\nu_j$ is
\begin{equation}
    \nu_j = \left|2\lambda_j -1\right|\,,
\end{equation}
and the bipartite entanglement entropy reads
\begin{equation}
    S_A=\sum_j \text{h}_2\left(\lambda_j\right),
\end{equation}
where, again, $\lambda_j$ are the eigenvalues of the covariance matrix restricted to $A$.

\subsection{The fermionic Choi, the OSEE, and the DW mapping}

Next, following Refs.~\cite{SunderhaufPiroliQi2019,FavaPiroliSwann2023,
DierckxFannesPogorzelska2008}, we introduce the Choi state. We first associate with the $L$ Majorana operators a label $a$, and introduce a second species of Majorana labeled by $b$, such that
\begin{equation}
    \{\psi_j^\alpha,\psi_k^\beta\}
    =
    2\delta_{jk}\delta_{\alpha\beta},
    \qquad
    \alpha,\beta\in\{a,b\}\,.
    \label{eq:clifford algebra global}
\end{equation}
The new Majorana modes allow us to double the degrees of freedom and introduce the Choi state. To this end, define the normalized complex modes
\begin{equation}
    \widetilde c_j
    =
    \frac{1}{2}
    \left(
        \psi_j^a-\ii\psi_j^b
    \right),
    \qquad
    \widetilde c_j^\dagger
    =
    \frac{1}{2}
    \left(
        \psi_j^a+\ii\psi_j^b
    \right),
    \label{eq:choi_dirac_modes}
\end{equation}
satisfying
\begin{equation}
    \{\widetilde c_j,\widetilde c_k^\dagger\}
    =
    \delta_{jk}.
\end{equation}
We introduce the maximally entangled state as their vacuum
\begin{equation}
    \widetilde c_j|I^{ab}\rangle\!\rangle=0.
\end{equation}
The Choi state of an operator $O$ is $|O\rangle\!\rangle = O^{(a)} |I^{ab}\rangle\!\rangle$, where the upper index $a$ specifies that it has a non-trivial expansion over the physical modes $a$ only.

Consider an unentangled Majorana string of arbitrary even length $M$
\begin{equation}
    O_M
    =
    \psi_1\psi_2\dots\psi_M\,,
\end{equation}
with Choi state
\begin{equation}
    |O_M\rangle\!\rangle
    =
    \psi_1^a\dots\psi_M^a|I^{ab}\rangle\!\rangle
    \propto
    \widetilde c_1^\dagger\dots
    \widetilde c_M^\dagger
    |I^{ab}\rangle\!\rangle.
    \label{eq:string_slater}
\end{equation}
Crucially,~\eqref{eq:string_slater} is a gaussian state.
To see this, consider the following identity for a pair of Majorana operators of the string
\begin{equation}
    \psi_j\,\psi_{j-1}=\exp\left({\frac{\pi}{2}\psi_j^a\psi_{j-1}^a}\right).
\end{equation}
Since all disjoint pairs commute, the initial string is rewritten as an exponential involving only quadratic Majorana operators
\begin{equation}
    O_M= \exp\left(\frac{\pi}{2}\sum_{j=1}^{M/2} \psi_{2j-1}\,\psi_{2j}\right),
    \label{eq:string is Gaussian operator}
\end{equation}
hence a gaussian unitary operator. Since a Fock vacuum evolved under a quadratic Hamiltonian is gaussian, the Choi state in~\eqref{eq:string_slater} is gaussian. This observation will allow us to perform both analytic and efficient numerical computations. We note that $|O_M\rangle\!\rangle$ is, in fact, a number conserving state with covariance matrix in the form of Eq.~\eqref{eq:covariance_matrixU1} given by a projector onto the $M$ modes of the string,
\begin{equation}
    G=P_M
    =
    \begin{pmatrix}
        \mathbf{1}_M & 0\\
        0 & 0_{L-M}
    \end{pmatrix}.
    \label{eq:initial_projector}
\end{equation}

In this work, we will be interested in the operator $O_M$ evolved under random gaussian evolution (cf. Sec.~\ref{sec:the_model}). Denoting by $O_M(t)$ the operator evolved up to time $t$, we have that $|O_M(t)\rangle\!\rangle$ is also gaussian. In order to compute its covariance matrix, we use standard techniques of fermionic gaussian states~\cite{surace2022fermionic}. In particular, given any quadratic operator $U_G=\exp (-i \psi^TH\psi)$, with $\psi=(\psi_1,\ldots,\psi_L ),$ we can associate a single-particle orthogonal operator $R\in {\rm SO}(L)$ such that
\begin{equation}
    U_G\psi_jU_G^\dagger
    =
    \sum_{k=1}^{L}{ R_{kj}}\psi_k,
    \qquad
    R\in\SO(L)\,.
    \label{eq:gaussian_majorana_rotation_intro}
\end{equation}
Therefore, $U_G$ maps a Majorana string into a coherent superposition of many
strings of the same weight. Note that, if $U_G$ is a generic gaussian unitary, it will mix modes lying on opposite
sides of a spatial bipartition, generating entanglement. Eq.~\eqref{eq:gaussian_majorana_rotation_intro} then also leads to the transformation of the covariance matrix
\begin{equation}\label{eq:transformation_equation}
   G \mapsto R  G R^T\,.
\end{equation}

The operator-space entanglement entropy (OSEE) is defined as the bipartite entanglement entropy of the Choi state. For gaussian states, we can use the same formalism described in Sec.~\ref{sec:Gaussian states and entanglement entropy}, yielding
\begin{align}
    \label{eq:OSEE}
    S_{\text{op}}(\ell)&=\sum_j \text{h}_2\left(\frac{1+\nu_j}{2}\right)=\sum_j \text{h}_2\left(\lambda_j\right) ,
\end{align}
where
\begin{equation}
    \text{h}_2(x)= -x\log x -(1-x)\log(1-x),
\end{equation}
while $\lambda_j$ are the eigenvalues of the Choi state covariance matrix $ G_A$, restricted to the region $A$ defining the bipartition. Importantly, we will consider a bipartition of the system such that $A$ contains the first $\ell=2N_A$ Majoranas. The string is also chosen to occupy the first $M$ Majorana labels; consequently one of these two sets contains the other. 

Eqs.~\eqref{eq:initial_projector}, ~\eqref{eq:transformation_equation} and~\eqref{eq:OSEE} completely encode the dynamics of the OSEE. Crucially, we see from these equations that the problem is now mapped onto that of a quantum quench from a DW initial state. Indeed, the dynamics is formally identical to that studied in Ref.~\cite{BernardPiroliScopa2026}, analyzing the entanglement entropy evolution of an initial DW state, with one important difference. That is, the covariance matrix evolves under the rule~\eqref{eq:transformation_equation} where $R$ is an orthogonal matrix taken from the group $\SO(L)$, while the evolution in Ref.~\cite{BernardPiroliScopa2026} was defined in terms of a unitary update $G\mapsto R G R^\dagger$, with $R$ a unitary matrix. In the rest of this work we will explore the consequences of this difference.

\subsection{The model}
\label{sec:the_model}

We will focus on a model of random gaussian dynamics known as the Brownian SYK$_2$ model~\cite{SunderhaufPiroliQi2019,jian2021note,agarwal2022emergent,BernardPiroliScopa2026}. The dynamics is defined by a time-dependent two-local Hamiltonian of the form
\begin{equation}
     H(t)= \frac{i}{2}\sum_{i<j} J_{ij}(t) \psi_i \psi_j\,,
     \label{eq:HamSYK2}
\end{equation}
where, $J_{ij}(t)$ are random parameters that are sampled independently for any pair $i,j$ from a suitable distribution. More precisely, consider dividing the interval $[0,t]$ into time steps $\Delta t$. During each one of these intervals, denoting $\Delta t=t_{n+1}-t_n$, we let the system evolve under the Hamiltonian $H(t_n)$, while we choose
\begin{equation}\label{eq:distributions}
    \EE [J_{ij}(t_n)]= 0, \quad  \EE [J_{ij}(t_n)J_{i'j'}(t_m)]= \frac{\delta_{i,i'}\delta_{j,j'}\delta_{n,m}}{\Delta t L}\,.
\end{equation}
The Brownian SYK$_2$ dynamics is obtained by taking $\Delta t\to 0$. In this limit, the evolution defines a stochastic process, which is best described by means of It\^o calculus~\cite{oksendal2013stochastic}. In this formalism, the dynamics is defined by the Hamiltonian generator
\begin{equation}
     {\rm d}H(t)= \frac{i}{2}\sum_{i<j} {\rm d}K_{ij}(t) \psi_i \psi_j\,,
     \label{eq:HamSYK2_ito}
\end{equation}
where $\dd K_t^T=-\dd K_t$ are real, antisymmetric Brownian increments satisfying
\begin{equation}
    \dd K_{ij}\dd K_{kl}
    =
    \frac{\dd t}{L}
    \left(
        \delta_{ik}\delta_{jl}
        -
        \delta_{il}\delta_{jk}
    \right)\,.
    \label{eq:antisymmetric_covariance}
\end{equation}

Note that, given an initial operator $O_M$, the stochastic dynamics gives rise to an ensemble of \emph{quantum trajectories}, each corresponding to one particular history of the random increments ${\rm d}K_{ij}(t)$. Along each trajectory the dynamics remains gaussian, and we can describe it using the covariance matrix formalism. That is, for each time $t$ there is a matrix $R_t\in{\rm SO}(L)$ such that the covariance matrix of the evolved state is
\begin{equation}
    G(t)=R_t G(0)R_t^{T}\,.
\end{equation}
In fact, following Ref.~\cite{BernardPiroliScopa2026}, we can write down the stochastic equation satisfied by $R_t$, which is
\begin{equation}
    \dd R_t
    =
    \dd K_t\,R_t
    -
    \alpha_LR_t\,\dd t,
    \qquad
    \alpha_L
    =
    \frac{L-1}{2L}\,.
    \label{eq:orthogonal_brownian}
\end{equation}
Equivalently, the process is defined by the infinitesimal update rule
\begin{equation}
    R_{t+\dd t}
    =
    e^{\dd K_t}R_t.
\end{equation}
In the next sections we will study the dynamics~\eqref{eq:orthogonal_brownian}. After treating its late-time limit in Sec.~\eqref{sec:page_curves}, we will discuss its exact solution at finite times in Sec.~\ref{sec:OSEE_dynamics}.

\section{Haar-averaged Page curves for the OSEE}
\label{sec:page_curves}

When $t\to\infty$, the Brownian motion~\eqref{eq:orthogonal_brownian} converges to the Haar random distribution over the space ${\rm SO}(L)$, and the analysis of the average OSEE simplifies. In this section, we will make use of the so-called Coulomb-gas approach, which is a standard tool in RMT~\cite{forrester2010log}. Our analysis will follow very closely Ref.~\cite{BernardPiroli2021}.

First, recall that a fermionic gaussian unitary acts on the original Majoranas through the
orthogonal rotation of Eq.~\eqref{eq:gaussian_majorana_rotation_intro}.
Because both Choi species undergo the same rotation, the
fermions $\tilde{c}_j$ transform as
\begin{equation}
    \widetilde c_j
    \longmapsto
    \sum_{k=1}^L R_{kj}\widetilde c_k.
\end{equation}
Consequently,
\begin{equation}
    G
    =
    RP_MR^T.
    \label{eq:rotated_projector}
\end{equation}
Let $A$ contain the first $\ell$ one-particle labels and write the first
$M$ columns of $R$ as
\begin{equation}
    \begin{pmatrix}
        X\\
        Y
    \end{pmatrix},
    \qquad
    X\in\mathbb{R}^{\ell\times M},
    \qquad
    Y\in\mathbb{R}^{(L-\ell)\times M}.
    \label{eq:X_Y_blocks}
\end{equation}
The restricted correlation matrix is
\begin{equation}
    G_A
    =
    XX^T.
    \label{eq:GA_XXT}
\end{equation}
The nonzero eigenvalues of $G_A$ coincide with those of
\begin{equation}
    J
    =
    X^TX.
    \label{eq:J_XTX}
\end{equation}
Let $R$ be Haar distributed in $\SO(L)$. Then $X$ is a rectangular block of a Haar-orthogonal matrix and $G_A=XX^T$ belongs to the real Jacobi ensemble~\cite[Thm.~1.5]{edelmansutton2008}, with Dyson index~\cite{Demni2008,DemniHamdiHmidi2012} 
\begin{equation}
    \beta=1 .
\end{equation}
In the region
\begin{equation}
\label{eq: region}
    \ell \leq M, \quad \ell +M \leq L , 
\end{equation}
all stationary $\ell$ eigenvalues fluctuate in the open interval $(0,1)$ and their joint density is~\cite{forrester2010log}
\begin{equation}
    \label{eq:joint density}
    \mathbb{P}(\lambda_1, \dots,\lambda_\ell) \propto \prod_{i<j} \left| \lambda_i - \lambda_j\right|^\beta\prod_{i=1}^\ell \lambda_i^{\frac{\beta}{2}(M-\ell +1)-1}\left(1-\lambda_i\right)^{\frac{\beta }{2}(L-M-\ell +1)-1} .
\end{equation}
In this work, we focus on studying the probability distribution for the orthogonal ensemble $\beta=1$. Note that, conversely, the unitary ensemble $\beta=2$ emerges when studying the stationary probability distribution obtained from the covariance matrix spectra of a random state undergoing two-local random unitary evolution~\cite{BernardPiroli2021}. 
Next, we take the thermodynamic limit, $L, \ell, M \to \infty$, with fixed ratios
\begin{equation}
    L,\ \ell,\ M \to \infty, \qquad q = \frac{\ell}{L}, \qquad \theta = \frac{M}{L}\,.
\end{equation}
In this limit, the discrete set of eigenvalues becomes a continuous density distribution function which we denote by $\rho_\star(\lambda)$. As detailed in ~\ref{app:coulomb_gas}, the stationary density is obtained using Coulomb-gas techniques developed in the random matrix literature~\cite{BernardPiroli2021}, yielding
\begin{equation}
    \rho_\star(\lambda) = \frac{1}{2\pi q}\,\frac{\sqrt{(\lambda_+ - \lambda)(\lambda - \lambda_-)}}{\lambda(1-\lambda)}, \qquad \lambda \in [\lambda_-, \lambda_+],
    \label{eq:wachter_density}
\end{equation}
with endpoints
\begin{equation}
\label{eq:wachter_endpoints}
    \lambda_\pm = \left(\sqrt{\theta(1-q)} \pm \sqrt{q(1-\theta)}\right)^2.
\end{equation}
% I commented this as in App B we justify h_2=0.{\color{red} Outside region \eqref{eq: region}, zero or one eigenvalues may occur at the hard edges
% \begin{equation}
%     n_0=\text{max}(0,\ell -M)\, , \quad n_1=\text{
%     max
%     }(0, \ell +M -L) ,
% \end{equation}
% but they do not contribute to the entropy since    $\text{h}_2(0)=\text{h}_2(1)=0.$

We note that in the thermodynamic limit, the orthogonal $\beta=1$ and $\beta=2$ ensembles yield the same eigenvalue density distribution~\cite{BernardPiroli2021,MurcianoCalabresePiroli2022,bianchi2022volume}.  
In the canonical region \eqref{eq: region} the continuous density is normalized
\begin{equation}
    \label{eq:sanity check 1}
    \int_{\lambda_-}^{\lambda_+} \rho_\star(\lambda),\, d\lambda = 1
\end{equation}
and its first moment is
\begin{equation}
    \label{eq:sanity check 2}
    \int_{\lambda_-}^{\lambda_+} \lambda\, \rho_\star(\lambda)\, d\lambda = \theta.
\end{equation}

\begin{figure}[t]
    \centering  \includegraphics[width=\textwidth,height=0.26\textheight]{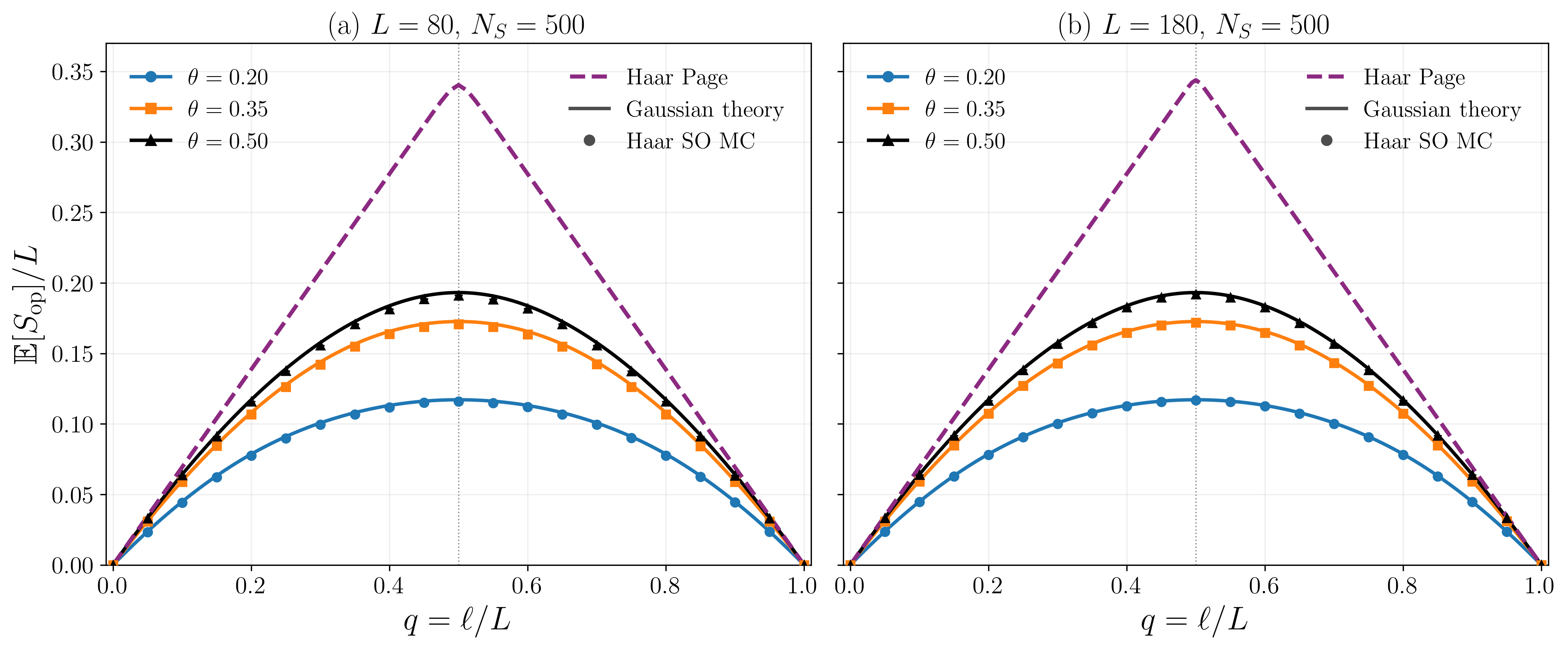}
    \caption{
        Stationary gaussian Page curves for $N_S=500$ Haar samples at $L=80$, left panel, and $L=180$ for the right panel.  The numerical Haar--$\SO(L)$
        data (dots) are compared with the Coulomb-gas prediction (bold curves)
        Eq.~\eqref{eq:page curve stationary}. 
        Both panels also show the
        fully Haar-random non-gaussian reference used in the numerical study,
        emphasizing that gaussian scrambling produces a distinct,
        $\theta$-dependent family of Page curves.
    }
    \label{fig:page_curve_placeholder}
\end{figure}

The expected operator entanglement entropy in the thermodynamic limit is obtained by integrating the binary entropy via a Coulomb-gas saddle point calculation, adapted from Ref.~\cite{BernardPiroli2021}. This yields the stationary gaussian Page curve
\begin{equation}
    \frac{\mathbb{E}[S_{\mathrm{op}}(\ell)]}{L} = q \int_{\lambda_-}^{\lambda_+} \rho_*(\lambda) h_2(\lambda) \, \mathrm{d}\lambda\,.
    \label{eq:page curve stationary}
\end{equation}
In Fig.~\ref{fig:page_curve_placeholder} we observe that the finite size OSEE curves approach the thermodynamic limit predictions for the stationary gaussian Page curves Eq.~\eqref{eq:page curve stationary} as we increase the system size $L$. It is also clear that the curves obtained for different filling fractions $\theta$ differ from the predictions obtained for fully Haar-random many-body states~\cite{Page1993}. 

\section{OSEE vs State Entanglement Dynamics}
\label{sec:OSEE_dynamics}

In this section, we analyze the dynamics of OSEE, by directly exploiting Eq.~\eqref{eq:orthogonal_brownian}. Details of the intermediate derivations are highlighted in~\ref{app:real_jacobi_derivation}.
In order to obtain the evolution of $G_A$, we decompose the first $M$ columns of the Brownian increment as
\begin{equation}
    \dd K_t
    =
    \begin{pmatrix}
        \dd A_t & \dd B_t\\
        -\dd B_t^T & \dd C_t
    \end{pmatrix}.
    \label{eq:K_block}
\end{equation}
This block representation yields the following update equation
\begin{equation}
    \dd X
    =
    \dd A\,X
    +
    \dd B\,Y
    -
    \alpha_LX\,\dd t,
    \label{eq:dX_real}
\end{equation}
where $X$ is defined in Eq.~\eqref{eq:X_Y_blocks}.
Using $X^TX+Y^TY=\mathbf{1}_M$,
and applying the It\^o product rule to $J=X^TX$, one obtains the closed
matrix Jacobi process ($\beta=1$) for $J$ \cite{BernardPiroliScopa2026,Demni2008,DemniHamdiHmidi2012}
\begin{equation}
    \begin{aligned}
    \dd J
    ={}&
    (\mathbf{1}-J)^{1/2}\dd W^T J^{1/2}
    +
    J^{1/2}\dd W(\mathbf{1}-J)^{1/2}
 +
    \left(
        q\mathbf{1}_M-J
    \right)\dd t,
    \end{aligned} 
    \label{eq:real_matrix_jacobi}
\end{equation}
where $\dd W$ is an $M\times M$ real Brownian matrix satisfying
\begin{equation}
    \dd W_{ij}\dd W_{kl}
    =
    \frac{\dd t}{L}
    \delta_{ik}\delta_{jl}.
    \label{eq:effective_real_brownian}
\end{equation}
Bringing the Jacobi matrix $J$ into diagonal form
\begin{equation}
    J(t)
    =
    O(t)\Lambda(t)O^T(t),
    \qquad
    \Lambda(t)
    =
    \operatorname{diag}
    \left(
        \lambda_1(t),\ldots,\lambda_M(t)
    \right),
\end{equation}
applying second-order perturbation theory and having in mind that the off-diagonal
stochastic increments are of order $\sqrt{\dd t}$, we arrive at the stochastic differential equation (SDE) for the eigengalues
\begin{equation}
    \begin{aligned}
    \dd\lambda_i
    ={}&
    \frac{2}{\sqrt{L}}
    \sqrt{
        \lambda_i(1-\lambda_i)
    }
    \,\dd\nu_i
    +
    \left[
        q-\lambda_i
        +
        \frac{1}{L}
        \sum_{j\neq i}
        \frac{
            \lambda_i(1-\lambda_j)
            +
            \lambda_j(1-\lambda_i)
        }{
            \lambda_i-\lambda_j
        }
    \right]\dd t,
    \end{aligned}
    \label{eq:real_jacobi_eigenvalues}
\end{equation}
where
\begin{equation}
    \dd\nu_i\dd\nu_j
    =
    \delta_{ij}\dd t.
\end{equation}
It is important to note that for the case of random gaussian evolution of an initial DW due to an all-to-all fermionic Hamiltonian, the complex process of Ref.~\cite{BernardPiroliScopa2026} has the same drift term but a noise coefficient $\sqrt{2/L}$ instead of $2/\sqrt{L}$.
These correspond to the Jacobi $\beta=2$ (unitary) and $\beta=1$ (orthogonal) processes, respectively. 
Next, we introduce the normalized empirical moments
\begin{equation}
    p_n^{(L)}(t)
    =
    \frac{1}{M}
    \Tr[J(t)^n]
    =
    \frac{1}{M}
    \sum_{i=1}^{M}
    \lambda_i(t)^n,
    \qquad
    p_0^{(L)}(t)=1.
    \label{eq:finite_L_empirical_moments}
\end{equation}
Taking their expectation value 
\begin{equation}
    m_n(t)
    =
    \EE[p_n^{(L)}(t)],
    \qquad
    m_0(t)=1,
    \label{eq:moment_definition}
\end{equation}
and using Eq~\eqref{eq:real_jacobi_eigenvalues}, we find
%\begin{equation}
%    \begin{aligned}
%     \partial_t m_n
%    &=
%    -n m_n
%    +
%    qn m_{n-1}
%    +
%    \theta n
%    \sum_{k=0}^{n-2}
%    m_{n-k-1}
%    \left(
%        m_k-m_{k+1}
%    \right)+
%    \frac{n(n-1)}{L} \left(  \frac{2}{\beta} -1\right)
%    \left[
%        m_{n-1}-m_n
%    \right].
%    \end{aligned}
%\end{equation}

    \begin{align}
     \partial_t m_n
    &{=
    -n m_n
    +
    qn m_{n-1}
    +
    \theta n\mathbb{E}\left[
    \sum_{k=0}^{n-2}
    p^{(L)}_{n-k-1}
    \left(
        p^{(L)}_k-p^{(L)}_{k+1}
    \right)
    \right]}\nonumber\\
    &{+
    \frac{n(n-1)}{L} \left(  \frac{2}{\beta} -1\right)
    \left[
        m_{n-1}-m_n
    \right].}
    \label{eq:finite_L_expected_moments}
    \end{align}
Note that the third term on the right-hand side of the expression, which involves expectations values of different moments, do not factorize at finite system sizes $\EE[p_a p_b]= m_a m_b + \text{Cov}(p_a,p_b)$. Eq.~\eqref{eq:finite_L_expected_moments} makes the $\beta$-dependence particularly transparent, although is it subleading in $1/L$.  

Indeed, at leading
thermodynamic order, Eq.~\eqref{eq:finite_L_expected_moments} becomes \cite{Demni2008,DemniHamdiHmidi2012,BernardPiroliScopa2026}
\begin{equation}
    \partial_t\mu_n
    =
    -n\mu_n
    +
    qn\mu_{n-1}
    +
    \theta n
    \sum_{k=0}^{n-2}
    \mu_{n-k-1}
    \left(
        \mu_k-\mu_{k+1}
    \right),
\label{eq:moment_hierarchy_unified}
\end{equation}
where we introduced
\begin{equation}
    \mu_n(t)
    =
    \lim_{\substack{L,\ell,M\rightarrow\infty\\
                    q,\theta\ \mathrm{fixed}}}
    \frac{1}{M}
    \EE\Tr[J(t)^n],
    \qquad
    \mu_0(t)=1.
    \label{eq:ThLimit_moment_definition}
\end{equation}
Introducing the resolvent for the eigenvalue distribution $\{ \boldsymbol{\lambda }\}$
\begin{equation}
    \mathcal{G}(z,t)
    =
    \int_0^1
    \frac{\rho_t(\lambda)}{z-\lambda}
    \,\dd\lambda
    =
    \frac{1}{z}
    \sum_{n=0}^{\infty}
    \frac{\mu_n(t)}{z^n},
    \label{eq:resolvent_definition}
\end{equation}
Eq.~\eqref{eq:moment_hierarchy_unified} is equivalent to the compact nonlinear
partial differential equation (PDE)
\begin{equation}
    \partial_t\mathcal{G}
    =
    \partial_z
    \left\{
        \left[
            (1-2\theta)z-(q-\theta)
        \right]\mathcal{G}
        +
        \theta z(z-1)\mathcal{G}^2
    \right\}.
    \label{eq:resolvent_pde}
\end{equation}
The detailed mathematical derivation of
Eqs.~\eqref{eq:moment_hierarchy_unified} and
\eqref{eq:resolvent_pde}, and their solutions are outlined in
Refs.~\cite{Demni2008,DemniHamdiHmidi2012}, but here we provide an alternative solution for this nonlinear PDE by introducing a Hamiltonian characteristic flow which decouples Eq\eqref{eq:resolvent_pde} by means of a characteristic energy $\mathcal{E}$. Starting with a balanced DW,
\begin{equation}
    M=\ell=\frac{L}{2},
    \qquad
    q=\theta=\frac{1}{2},
    \label{eq:balanced_domain_wall}
\end{equation}
the PDE simplifies to
\begin{equation}
    \partial_t\mathcal{G}
    =
    \frac{1}{2}
    \partial_z
    \left[
        z(z-1)\mathcal{G}^2
    \right],
\label{eq:domain_wall_resolvent_pde}
\end{equation}
with initial conditions 
\begin{equation}
    J(0)=\mathbf{1}_M,
    \qquad
    \mu_n(0)=1,
    \qquad
    \mathcal{G}(z,0)=\frac{1}{z-1}.
    \label{eq:domain_wall_initial_condition}
\end{equation}
The stationary solution is the arcsine distribution.  This allows us to compute the moments based on the method of characteristics, cf.~\ref{app:characteristics}, obtaining a solution that coincides with the one established in
Refs.~\cite{Demni2008,DemniHamdiHmidi2012,BernardPiroliScopa2026}. The moments read
\begin{equation}
    \mu_n(t)
    =
    \frac{1}{4^n}\binom{2n}{n}
    +
    \frac{2}{4^n}
    \sum_{k=1}^{n}
    \binom{2n}{n-k}
    \frac{L_{k-1}^{(1)}(2kt)}{k}e^{-kt}.
    \label{eq:moment_balanced}
\end{equation}
Here $L^{(1)}_{n}$ denotes the Laguerre polynomial of index one. Given Eq.\eqref{eq:moment_balanced}, the thermodynamic OSEE density coincides with the spectral
functional as the von-Neumann entropy density considered by Ref.~\cite{BernardPiroliScopa2026},
\begin{equation}
    s_{\rm op}(t)
    :=\lim_{L\to\infty}\frac{\EE[\OSEE(t)]}{\ell}
    =\int_0^1\rho_t(\lambda)h_2(\lambda)\,\dd\lambda.
    \label{eq:entropy_density_from_spectrum}
\end{equation}
Using the convergent expansion of $h_2$ around $1/2$ and inserting
Eq.~\eqref{eq:moment_balanced}, one obtains Eq.~(46) of
Ref.~\cite{BernardPiroliScopa2026}, which in our time convention reads
\begin{equation}
\begin{aligned}
    s_{\rm op}(t)
    ={}&2\log2-1
    -\sum_{n=1}^{\infty}\frac{1}{n(2n-1)}
      \sum_{k=1}^{n}
      \frac{(-2)^{-2k}}{2k}
      \binom{2n}{2k}
      \\
    &\times
      {}_2F_1\!\left(
          2k+\frac12,\,2k-2n;\,4k+1;\,2
      \right)
      e^{-2kt}L_{2k-1}^{(1)}(4kt).
\end{aligned}
\label{eq:scopa_entanglement_formula}
\end{equation}
In particular, all Laguerre contributions vanish at long times and
$s_{\rm op}(\infty)=2\log2-1$, consistently with the stationary arcsine law. This expression can be further simplified by parameterizing the integrand of Eq.~\eqref{eq:entropy_density_from_spectrum} so that $\lambda= \cos^2(\phi)/2$. Expanding $\lambda^n= (e^{i\phi/2}+ e^{-i\phi/2})^n/2^n$ and matching coefficients with Eq.~\eqref{eq:moment_balanced} we find that 
\begin{equation}
    c_n(t)= \int^1_0 d\lambda \rho_t(\lambda) \cos(n\phi)= (L^{(1)}_{n-1}(2nt) e^{-nt})/n.
    \label{eq:c_n}
\end{equation}
Finally, rewriting the binary entropy density $\rm h_2(\lambda)$ in terms of trigonometric functions using $\log(2\sin(\phi/2) )= -\sum_m \cos^m(\phi)/m, \quad \log(2\cos(\phi/2) )=\sum_m (-1)^{m-1}\cos^m(\phi)/m$, and integrating over the spectrum density, we obtain
\begin{equation}\label{eq:simplified_eq}
    s_{\rm op}(t)=2\log 2-1-\sum_{k=1}^{\infty} \frac{e^{-2kt}L_{2k-1}^{(1)}(4kt)}{2k^2(4k^2-1)}\,.
\end{equation}

% ============================================================
% General subsystem and filling fractions
% ============================================================
Before leaving this section, we report an extension of the characteristic approach developed above beyond the
balanced point $q=\theta=1/2$. Starting from the thermodynamic resolvent equation
\begin{equation}
    \partial_t \cG(t,z)
    =
    \partial_z
    \left\{
    \left[(1-2\theta)z-(q-\theta)\right]\cG
    +
    \theta z(z-1)\cG^2
    \right\},
    \label{eq:general_resolvent_pde}
\end{equation}
we show in~\ref{app:general_characteristics} that the associated
characteristic flow is Hamiltonian and admits a conserved energy. We note that finite-time moments of the free Jacobi process for arbitrary parameters have
recently been obtained for
general coefficients, by means of finite recurrence
relations~\cite{DemniHamdi2025Moments}. Our solution is complementary to previous methods, allowing us to analytically track the OSEE dynamics without the need to perform sums over the distribution moments. Introducing a characteristic parameter $s$ and restricting first to
$q\geq\theta$, the full thermodynamic solution can be written parametrically
in terms of an auxiliary function
\begin{equation}
    h(s,t)
    =
    s\,
    \frac{1+s+(1-s)e^{st}}
         {1+s-(1-s)e^{st}}.
    \label{eq:h_general_main}
\end{equation}
Defining the constants
\begin{equation}
    a=1-2\theta,
    \qquad
    b=2q-1,
\end{equation}
the spectral coordinate and resolvent obey the compact relations
\begin{equation}
    z(s,t)
    =
    1-\frac{(a-h)(b-h)}{h^2-s^2},
    \qquad
    (z-1)\cG
    =
    \frac{h-a}{2\theta}.
    \label{eq:general_parametric_solution_main}
\end{equation}
The complementary regime $q<\theta$ follows from the equality of the
nonzero spectra of $X^TX$ and $XX^T$.
Importantly, although there is no evident simpler solution for the thermodynamic normalized moments in terms of special functions, the OSEE can be evaluated
directly from the resolvent without reconstructing the time-dependent
spectral density. Using an integral representation of the binary entropy,
we obtain
\begin{equation}
    \frac{S_{\rm op}(t)}{L}
    =
    \theta
    \int_0^\infty d\tau\,
    \left[
        2-\frac{1}{1+\tau}
        +
        \tau\left(
            \cG(t,-\tau)-\cG(t,1+\tau)
        \right)
    \right].
\label{eq:general_entropy_resolvent}
\end{equation}
Thus, the thermodynamic OSEE dynamics at arbitrary $q$ and $\theta$ are
reduced to a one-dimensional real quadrature of the exact parametric
resolvent. At $q=\theta=1/2$, Eqs.~\eqref{eq:h_general_main} and
\eqref{eq:general_parametric_solution_main} reduce exactly to the conformal
map and resolvent obtained in the balanced calculation.

\begin{figure}[ht]
    \centering
    \includegraphics[width=\linewidth,height=0.33\textheight]{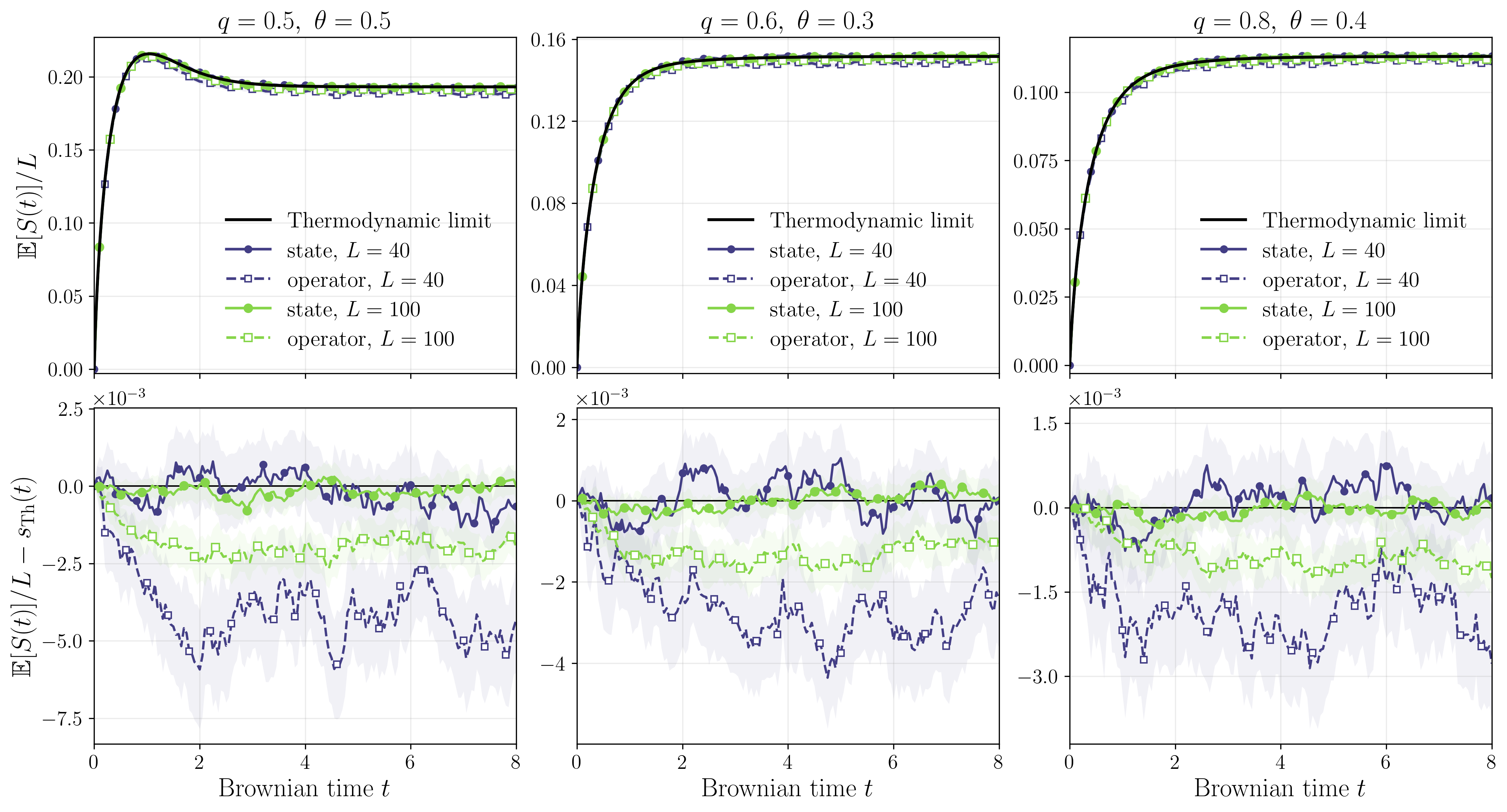}
    \caption{
        Numerical comparison of the real ($\beta=1$) operator Jacobi dynamics and the
        complex ($\beta=2$) state Jacobi dynamics with the common thermodynamic limit,
        for the balanced bipartition $q=\theta=1/2$ (left), $q=0.6, \; \theta=0.3$ (middle), $q=0.8, \; \theta=0.4$ (right).  Circles denote the state entanglement
        obtained from complex random-unitary evolution of a DW state, while squares
        denote the OSEE generated by the gaussian evolution of a Majorana string.
        Lower row: finite size deviations of the entropy-density dynamics; the black curve is the thermodynamic prediction
Eqs.~(\ref{eq:scopa_entanglement_formula}, \ref{eq:general_entropy_resolvent}) .} 
    \label{fig:entanglement_finite_size}
\end{figure}

In
Fig.~\ref{fig:entanglement_finite_size} we compare the OSEE (corresponding to $\beta=1$ in Eq.~\eqref{eq:finite_L_expected_moments}) with the state-entanglement dynamics (corresponding to $\beta=1$). We see that the two curves are the same in the thermodynamic limit, but the finite-size corrections are different.  The $\beta=2$ data
converge rapidly to the thermodynamic curve, whereas the OSEE ($\beta=2$) retain a
visible systematic correction even at the largest system size shown.  This is precisely the pattern
expected from the finite-$L$ moment equation: the explicit correction is proportional to
$(2/\beta-1)/L$, so it cancels for $\beta=2$ but survives for $\beta=1$; see
Eq.~\eqref{eq:finite_L_expected_moments}.  The lower panels show the difference between the two curves, which decreases as $L$ increases.

\section{Implications for operator-space non-gaussianity}
\label{sec:non_gaussianity}

In this section, we discuss the implications of our results on the possibility to put an upper bound on the OSEE in terms of the operator-space non-gaussianity. 
In Ref.~\cite{DowlingModiWhite2025}, the authors derived an upper bound on the OSEE in terms of operator-space nonstabilizerness, taking the form
\begin{equation}
    S^{(\alpha)}_{\rm op}(O)\leq M_{\alpha}(O)\,, 
\end{equation}
where $M_{\alpha}(O)$ is the operator-space stabilizer R\'enyi entropy~\cite{DowlingKosTurkeshi2025}. This inequality is interesting because it directly relates two notions of operator complexity. As our work studies the growth of OSEE under gaussian dynamics, one could wonder whether a similar relation holds for a suitably defined measure of non-gaussianity in operator space.

We inspect a bound of the form
\begin{equation}\label{eq:conjecture}
    S_{\rm op}(O)\leq c \mathcal{F}(O)\,,
\end{equation}
for some constant independent of the system size, and take $\mathcal{F}(O)$ to be some measure of non-gaussianity in operator space. Choosing $O=U \psi_{1}\ldots \psi_{N} U^\dagger$, where $U$ is a random gaussian unitary, we obtain from Sec.~\ref{sec:page_curves} that $S_{\rm op}(O)\propto L$. Therefore, it must also be $\mathcal{F}(O)=\Omega (L)$. At the same time, if $\mathcal{F}$ is a meaningful measure of non-gaussianity, it must be $\mathcal{F}(O)=\mathcal{F}(\psi_{1}\ldots \psi_{N})$, and therefore
\begin{equation}\label{eq:identification}
    \mathcal{F}(\psi_{1}\ldots \psi_{N})=\Omega(L)\,.
\end{equation}

From the above discussion it is clear that $\mathcal{F}(O)$ cannot be taken to be the non-gaussianity of the vectorized operator $|O\rangle\!\rangle$, since we showed in Sec.~\ref{sec:Gaussian states and entanglement entropy} that $\psi_{1}\ldots \psi_{N}$ is a gaussian operator. Rather, Eq.~\eqref{eq:identification} suggests defining $\mathcal{F}$ as a quantity which is extensive in the number of Majoranas in the operator support. A natural candidate would be the total Majorana weight introduced in Ref.~\cite{lyu2024fermionic}. Given an expansion of an operator $O$ in terms of Majorana strings $\Psi_S$, $O=\sum_S a_S\Psi_S$, the total Majorana weight reads 
\begin{equation}
    W(O)=\sum_S |S|\,|a_S|^2,
\end{equation}
where $\sum_S|a_S|^2=1$, while $|S|$ is the number of Majorana forming $\Psi_S$. Note that this is not a faithful monotone, as gaussian operators have non-zero weight.

For the Majorana string $O=\psi_{1}\ldots \psi_{N}$, we have $W(O)=N$. However, consider
\begin{equation}
O_m=\frac1{\sqrt m}\sum_{j=1}^{m}
\psi_{2j-1}\psi_{2j}\psi_{2m+1}\psi_{2m+1+j},
\end{equation}
with $L=4m$ and $A=\{1,\ldots,2m\}$. $O_m$ is a unitary Hermitian operator, with $W(O_m)=4$, but the OSEE is $\log m$, so that~\eqref{eq:conjecture} is not satisfied. These considerations rule out using the total Majorana weight as a useful measure of operator non-gaussianity. Although different measures are possible, this discussion highlights the difficulty to extend the inequality of Ref.~\cite{DowlingModiWhite2025} to non-gaussianity, without any additional term that may depend on the operator to be evolved.

\section{Conclusions}
\label{sec:conclusions}
In this work, we have studied the OSEE of random fermionic gaussian dynamics, leveraging tools from quantum many-body physics~\cite{
ProsenPizorn2007,
Zanardi2001,
Dubail2017,
ZhouLuitz2017,
WangZhou2019,
bertini2020operator,
ZhouNahum2020,
KudlerFlamEtAl2021} and RMT~\cite{BernardPiroliScopa2026,Demni2008,DemniHamdiHmidi2012,demni2025spectral,demni2025finite}. By putting an arbitrary Majorana string in an enlarged space we identified a mapping between the Choi-state representation of the operator and that of DW states. Using this map, the computation of OSEE was reduced to studying the spectrum of the Choi covariance matrix, allowing us to obtain exact predictions for both the dynamics and steady state values of OSEE in the thermodynamic limit. Interestingly, the difference between random-state and random-operator entanglement dynamics lies on finite size  corrections, stemming from the different underlying group structure of the Jacobi processes. 

Our work raises several questions. Prominently, it would be interesting to obtain an alternative derivation of our results using replica calculation techniques~\cite{ProsenPizorn2007,Zanardi2001,
Dubail2017,
ZhouLuitz2017,
WangZhou2019,
bertini2020operator,
ZhouNahum2020,KudlerFlamEtAl2021}. Another direction pertains to the extension of our approach to different quantities, including the out-of-time order correlators (OTOCs)~\cite{maldacena2016remarks}. We leave these questions for future work. 

\section*{Note added}While finalizing this work, we became aware of Refs.~\cite{white2026exact} and~\cite{tonetto2026free} that also investigates the OSEE in random gaussian dynamics. The setting of Ref.~\cite{white2026exact} is different because the authors investigate global Haar-random matchgate circuits and compute averaged operator purity, corresponding to annealed R\'enyi-2 entropy. The setting of Ref.~\cite{tonetto2026free} is also different, as it investigates local random gaussian dynamics.

\section*{AI use disclosure}ChatGPT suggested the integral representation in Eq.~\eqref{eq:log_integral_rep} which facilitated the use of the general parametric solution of the Jacobi process resolvent to compute the bipartite entanglement dynamics. It also provided the simplification~\eqref{eq:simplified_eq}. The draft was written by the authors, who take full responsibility for the final manuscript and its scientific content.

\section*{Acknowledgments} We thank Neil Dowling and Federico Tonetto for useful discussions. This work was funded by the European Union (ERC, QUANTHEM, 101114881). Views and opinions expressed are however those of the author(s) only and do not necessarily reflect those of the European Union or the European Research Council Executive Agency. Neither the European Union nor the granting authority can be held responsible for them. 

%\newpage

% --------------------------------------------x----------------
%  Bibliography                    

%\printbibliography

% ============================================================
% Appendices
% ============================================================

\appendix

\section{Coulomb-gas derivation of the stationary density}
\label{app:coulomb_gas}
In this appendix we summarize the main steps leading from the finite-size Jacobi ensemble to
Eq.~\eqref{eq:wachter_density}. To obtain the stationary density we assume the filling fraction and occupation rate satisfy $q\leq \theta\; \text{and} \; q+\theta\leq 1$.
First, we introduce the empirical density
\begin{equation}
    \rho(\lambda)
    =
    \frac{1}{\ell}
    \sum_{j=1}^{\ell}
    \delta(\lambda-\lambda_j),
    \qquad
    \int_0^1\rho(\lambda)\,\dd\lambda=1.
\end{equation}
At leading order in $\ell$, the probability is governed by the energy
functional
\begin{equation}
    \begin{aligned}
    \mathcal{E}[\rho]
    ={}&
    -\frac{\beta}{2}
    \int\!\!\int
    \rho(\lambda)\rho(\mu)
    \log|\lambda-\mu|
    \,\dd\lambda\,\dd\mu + w \int d\lambda  \rho(\lambda) {\rm h }_2(\lambda)
    \\
    &+
    \frac{\beta}{2}
    \int
    \rho(\lambda)V(\lambda)\,\dd\lambda
    +
    u
    \left(
        \int\rho(\lambda)\,\dd\lambda-1
    \right),
    \end{aligned}
    \label{eq:coulomb_energy_appendix}
\end{equation}
with the Lagrange multiplier $u$ enforcing normalization of the distribution,  and
\begin{equation}
    V(\lambda)
    =
    -
    \left(
        \frac{\theta}{q}-1
    \right)\log\lambda
    -
    \left(
        \frac{1-\theta}{q}-1
    \right)\log(1-\lambda).
    \label{eq:effective_potential_appendix}
\end{equation}

Here, $\beta=1$ and $\beta=2$ correspond to the orthogonal and complex ensemble distributions respectively.
To obtain the density, we minimize the energy functional with respect to $\rho$ and obtain the saddle-point equation
\begin{equation}
    \beta \int
    \rho_*(\mu)
    \log|\lambda-\mu|
    \dd\mu= \frac\beta2 V(\lambda) +w {\rm h}_2(\lambda) +u .
\end{equation}
With this notation the averaged entanglement entropy is given by 
\begin{equation}
    S(l)= l \partial_w \mathcal{E}[\rho_*]|_{w=0}.
\end{equation}
In order to obtain $\rho_*$, we focus on the case where $w=0$
and differentiate with respect to $\lambda$, which gives
\begin{equation}
    \PV
    \int_{\lambda_-}^{\lambda_+}
    \frac{\rho_\star(\mu)}{\lambda-\mu}
    \,\dd\mu
    =
    \frac{1}{2}V'(\lambda).
    \label{eq:singular_integral_appendix}
\end{equation}
Note that at this point the $\beta$-dependence of the energy functional becomes irrelevant. Therefore, both ensembles yield the same stationary distribution in the thermodynamic limit.
Define the resolvent
\begin{equation}
    \cG(z)
    =
    \int_{\lambda_-}^{\lambda_+}
    \frac{\rho_\star(\mu)}{z-\mu}
    \,\dd\mu,
    \qquad
    \cG(z)\sim\frac{1}{z}
    \quad
    (z\rightarrow\infty).
\end{equation}
Therefore, the boundary values satisfy
\begin{equation}
    \cG(\lambda\pm\ii0)
    =
    \frac{1}{2}V'(\lambda)
    \mp
    \ii\pi\rho_\star(\lambda).
\end{equation}
Letting
\begin{equation}
    W(z)
    =
    \cG(z)-\frac{1}{2}V'(z),
\end{equation}
we introduce
\begin{equation}
    \mathfrak{r}(z)
    =
    \sqrt{
        (z-\lambda_-)(z-\lambda_+)
    },
    \qquad
    \mathfrak{r}(z)\sim z
    \quad
    (z\rightarrow\infty),
\end{equation}
such that $W$ and $\mathfrak r$ change sign across the cut.  Their ratio is
therefore meromorphic,
\begin{equation}
    \Phi(z)
    =
    \frac{W(z)}{\mathfrak r(z)}
    =
    \frac{1}{2q}
    \left(
        \frac{1}{z-1}
        -
        \frac{1}{z}
    \right).
\end{equation}
Cancellation of the poles at $z=0,1$ and the large-$z$ normalization give
\begin{equation}
    \sqrt{\lambda_-\lambda_+}
    =
    \theta-q,
\end{equation}
\begin{equation}
    \sqrt{
        (1-\lambda_-)(1-\lambda_+)
    }
    =
    1-\theta-q.
\end{equation}
Solving these equations yields
Eq.~\eqref{eq:wachter_endpoints}.  Taking the discontinuity of $G$ across
the cut gives Eq.~\eqref{eq:wachter_density}.

\section{Derivation of the real matrix Jacobi process}
\label{app:real_jacobi_derivation}

In this appendix we provide details on the real matrix Jacobi process, which describes the random dynamics of the Choi covariance matrix spectra. The geometric infinitesimal update of the orthogonal rotation is given
\begin{equation}
    R_{t+\dd t}
    =
    e^{\dd K_t}R_t,
    \qquad
    \dd K_t^T=-\dd K_t.
\end{equation}
Expanding the exponential,
\begin{equation}
    e^{\dd K_t}
    =
    \mathbf{1}
    +
    \dd K_t
    +
    \frac{1}{2}\dd K_t^2
    +
    O(\dd t^{3/2}),
\end{equation}
and using
\begin{equation}
    \dd K_t^2
    =
    -\frac{L-1}{L}
    \mathbf{1}_L\dd t,
\end{equation}
one finds
\begin{equation}
    \dd R_t
    =
    \dd K_tR_t
    -
    \frac{L-1}{2L}R_t\dd t.
\end{equation}

To extract the Jacobi matrix note that from Eq.~\eqref{eq:K_block},
\begin{equation}
    \dd X
    =
    \dd A\,X
    +
    \dd B\,Y
    -
    \alpha_LX\dd t,
\end{equation}
and
\begin{equation}
    \dd X^T
    =
    -X^T\dd A
    +
    Y^T\dd B^T
    -
    \alpha_LX^T\dd t,
\end{equation}
where  $\alpha_L= \frac{L-1}{2L}$.
Here, the required contractions are
\begin{equation}
    \dd A^T\dd A
    =
    \frac{\ell-1}{L}
    \mathbf{1}_\ell\dd t,
\end{equation}
\begin{equation}
    \dd B^T\dd B
    =
    \frac{\ell}{L}
    \mathbf{1}_{L-\ell}\dd t,
\end{equation}
and the mixed contractions vanish. Applying
\begin{equation}
    \dd J
    =
    (\dd X^T)X
    +
    X^T\dd X
    +
    \dd X^T\dd X
\end{equation}
gives
\begin{equation}
    \dd J
    =
    Y^T\dd B^TX
    +
    X^T\dd BY
    +
    \left(
        q\mathbf{1}_M-J
    \right)\dd t.
    \label{eq:preclosed_J_appendix}
\end{equation}

To avoid inconsistent results, we proceed our analysis by working in the canonical parameter region
\begin{equation}
    M\leq \ell,
    \qquad
    M\leq L-\ell,
\end{equation}
or equivalently
\begin{equation}
    \theta\leq q,
    \qquad
    \theta\leq 1-q.
    \label{eq:canonical_jacobi_region}
\end{equation}
In this region, away from the boundary where eigenvalues can reach
a hard edge, both $X$ and $Y$ have full column rank and admit the polar
decompositions
\begin{equation}
    X=UJ^{1/2},
    \qquad
    Y=V(\mathbf{1}-J)^{1/2},
\end{equation}
with
\begin{equation}
    U^TU=V^TV=\mathbf{1}_M.
\end{equation}

Note that the rank conditions required by the above polar decomposition do not hold over the
entire $(q,\theta)$ parameter space.  However, the entropy problem possesses
two exact transformations that allow us to reduce any point to a canonical
region.  First, exchanging $q$ and $\theta$ leaves the nonzero Jacobi spectrum
unchanged, since $XX^T$ and $X^TX$ have the same nonzero eigenvalues.  Second,
the simultaneous transformation
\begin{equation}
    (q,\theta)
    \longmapsto
    (1-q,1-\theta)
\end{equation}
maps the problem to the complementary pair of subspaces.  This transformation
preserves the fluctuating Jacobi spectrum up to deterministic hard-edge modes, $\lambda=\{0,1\}$,
which do not contribute to the OSEE because
$h_2(0)=h_2(1)=0$.

Using these two transformations, any $(q,\theta)$ can be mapped to canonical
parameters
\begin{equation}
(q_{\rm c},\theta_{\rm c})
=(l_{\rm c}/L,M_{\rm c}/L
)=
\begin{cases}
\bigl(\max(q,\theta),\,\min(q,\theta)\bigr),
& q+\theta\leq 1,\\[2mm]
\bigl(1-\min(q,\theta),\,1-\max(q,\theta)\bigr),
& q+\theta>1.
\end{cases}
\label{eq:canonical_parameter_map}
\end{equation}
Therefore the following relation holds 
\begin{equation}
    0\leq \theta_{\rm c}
    \leq
    \min(q_{\rm c},1-q_{\rm c}),
\end{equation}
so that
\begin{equation}
    M_{\rm c}\leq \ell_{\rm c},
    \qquad
    M_{\rm c}\leq L-\ell_{\rm c}.
\end{equation}

Note that outside the canonical region, apparent rank issues corresponds to
deterministic hard-edge eigenvalues at $0$ or $1$.  These modes are harmless because they have a zero 
contribution to the entropy, $h_2(0)=h_2(1)=0$. 
Defining
\begin{equation}
    \dd W
    =
    U^T\dd BV,
\end{equation}
we have
\begin{equation}
    \dd W_{ij}\dd W_{kl}
    =
    \frac{\dd t}{L}
    \delta_{ik}\delta_{jl},
\end{equation}
and Eq.~\eqref{eq:preclosed_J_appendix} can be rewritten as
Eq.~\eqref{eq:real_matrix_jacobi}.
At a fixed time, one can choose to work in the instantaneous eigenbasis of $J$ such that 
\begin{equation}
    J
    =
    \operatorname{diag}
    (\lambda_1,\ldots,\lambda_M).
\end{equation}
Hence, the stochastic off-diagonal increment of Eq.~\eqref{eq:real_matrix_jacobi} yields
\begin{equation}
    \dd N_{ij}
    =
    \sqrt{1-\lambda_i}\sqrt{\lambda_j}
    \,\dd W_{ji}
    +
    \sqrt{\lambda_i}\sqrt{1-\lambda_j}
    \,\dd W_{ij},
\end{equation}
and the diagonal terms $i=j$
\begin{equation}
    \dd N_{ii}
    =
    \frac{2}{\sqrt L}
    \sqrt{\lambda_i(1-\lambda_i)}
    \,\dd\nu_i,
\end{equation}
with
\begin{equation}
    \dd\nu_i\dd\nu_j
    =
    \delta_{ij}\dd t.
\end{equation}

For $i\neq j$ the following relation holds
\begin{equation}
    \dd N_{ij}\dd N_{ji}
    =
    \frac{\dd t}{L}
    \left[
        \lambda_i(1-\lambda_j)
        +
        \lambda_j(1-\lambda_i)
    \right].
\end{equation}
Assuming $J$ has a nondegenerate spectrum, second-order perturbation theory gives
\begin{equation}
    \dd\lambda_i
    =
    \dd J_{ii}
    +
    \sum_{j\neq i}
    \frac{
        \dd J_{ij}\dd J_{ji}
    }{
        \lambda_i-\lambda_j
    }.
\end{equation}
The second term must be retained because
$\dd J_{ij}=O(\sqrt{\dd t})$.  Substituting the quadratic variation above
yields Eq.~\eqref{eq:real_jacobi_eigenvalues}.

Next, we derive the evolution equation for the moments of the real Jacobi
process $p^{(L)}_n(t)=
    \frac{1}{M}
    \sum_{i=1}^{M}
    \lambda_i(t)^n,$ and compress the resulting infinite hierarchy into a nonlinear PDE
for the resolvent $\cG$. The real Jacobi eigenvalues satisfy
\begin{equation}
    \begin{aligned}
    \dd\lambda_i
    ={}&
    \frac{2}{\sqrt{L}}
    \sqrt{\lambda_i(1-\lambda_i)}
    \,\dd\nu_i
  +
    \left[
        q-\lambda_i
        +
        \frac{1}{L}
        \sum_{j\neq i}
        \frac{
            \lambda_i(1-\lambda_j)
            +
            \lambda_j(1-\lambda_i)
        }{
            \lambda_i-\lambda_j
        }
    \right]\dd t,
    \end{aligned}
    \label{eq:appendix_eigenvalue_sde}
\end{equation}
where
\begin{equation}
    q=\frac{\ell}{L},
    \qquad
    \theta=\frac{M}{L}\,.
\end{equation}
Applying It\^o's formula to a single eigenvalue gives
\begin{equation}
    \dd(\lambda_i^n)
    =
    n\lambda_i^{n-1}\dd\lambda_i
    +
    \frac{n(n-1)}{2}
    \lambda_i^{n-2}
    (\dd\lambda_i)^2.
    \label{eq:ito_lambda_power}
\end{equation}
The quadratic variation is
\begin{equation}
    (\dd\lambda_i)^2
    =
    \frac{4}{L}
    \lambda_i(1-\lambda_i)\dd t,
    \label{eq:eigenvalue_quadratic_variation}
\end{equation}
and the corresponding It\^o correction is therefore
\begin{equation}
    \frac{2n(n-1)}{L}
    \left(
        \lambda_i^{n-1}-\lambda_i^n
    \right)\dd t.
    \label{eq:ito_correction_single}
\end{equation}

After summing over the eigenvalues and dividing by $M$, the stochastic
contribution is the increment
\begin{equation}
    \dd\mathfrak{N}_n^{(L)}
    =
    \frac{2n}{M\sqrt{L}}
    \sum_{i=1}^{M}
    \lambda_i^{n-1}
    \sqrt{\lambda_i(1-\lambda_i)}
    \,\dd\nu_i.
    \label{eq:moment_martingale}
\end{equation}
The one-body drift gives
\begin{equation}
    -np_n^{(L)}
    +
    qn p_{n-1}^{(L)}.
    \label{eq:one_body_moment_drift}
\end{equation}
The only nontrivial term is the pairwise eigenvalue interaction coming from the second order perturbation contribution.  Introducing
\begin{equation}
    A_{ij}
    =
    \lambda_i(1-\lambda_j)
    +
    \lambda_j(1-\lambda_i),
    \qquad
    A_{ij}=A_{ji}.
    \label{eq:Aij_definition}
\end{equation}
Its contribution to $\dd p_n^{(L)}$ is
\begin{equation}
    \mathcal{I}_n^{(L)}
    =
    \frac{n}{ML}
    \sum_{i\neq j}
    \lambda_i^{n-1}
    \frac{A_{ij}}{\lambda_i-\lambda_j}.
    \label{eq:raw_interaction_moment}
\end{equation}

The singularity of the above expression can be removed by the polynomial identity
\begin{equation}
    \frac{x^{n-1}-y^{n-1}}{x-y}
    =
    \sum_{k=0}^{n-2}
    x^{n-k-2}y^k,
    \label{eq:divided_difference_identity}
\end{equation}
and using
\begin{equation}
    A_{ij}
    =
    \lambda_i+\lambda_j-2\lambda_i\lambda_j,
\end{equation}
the symmetry under $i\leftrightarrow j$, and 
$M/L=\theta$, we obtain
\begin{equation}
    \begin{aligned}
    \mathcal{I}_n^{(L)}
    ={}&
    n\theta
    \sum_{k=0}^{n-2}
    p_{n-k-1}^{(L)}
    \left(
        p_k^{(L)}-p_{k+1}^{(L)}
    \right)
    -
    \frac{n(n-1)}{L}
    \left(
        p_{n-1}^{(L)}-p_n^{(L)}
    \right).
    \end{aligned}
\label{eq:interaction_moment_final}
\end{equation}
The first line survives in the thermodynamic limit.  The second line is a
finite-size correction originating from the exclusion of the diagonal
$i=j$ in the pair interaction. 

Define the limiting moments
\begin{equation}
    \mu_n(t)
    =
    \lim_{\substack{L,M,\ell\rightarrow\infty\\
                    M/L=\theta,\ \ell/L=q}}
    \EE[p_n^{(L)}(t)],
    \qquad
    \mu_0(t)=1.
    \label{eq:limiting_moments_appendix}
\end{equation}
In the free-Jacobi limit, normalized traces become asymptotically
factorized:
\begin{equation}
    \EE[
        p_a^{(L)}p_b^{(L)}
    ]
    \longrightarrow
    \mu_a\mu_b,
    \label{eq:asymptotic_factorization}
\end{equation}
where
we use the free-Jacobi limit established in
Refs.~\cite{Demni2008,DemniHamdiHmidi2012,BernardPiroliScopa2026}.  Therefore, the explicit $1/L$ correction vanishes, and
Eq.~\eqref{eq:finite_L_expected_moments} becomes the closed hierarchy
\begin{equation}
    \partial_t\mu_n
    =
    -n\mu_n
    +
    qn\mu_{n-1}
    +
    \theta n
    \sum_{k=0}^{n-2}
    \mu_{n-k-1}
    \left(
        \mu_k-\mu_{k+1}
    \right).
\label{eq:thermodynamic_moment_hierarchy_appendix}
\end{equation}
The first two equations are
\begin{equation}
    \partial_t\mu_1
    =
    -\mu_1+q,
    \label{eq:first_moment_equation}
\end{equation}
and
\begin{equation}
    \partial_t\mu_2
    =
    -2\mu_2
    +
    2q\mu_1
    +
    2\theta\mu_1(1-\mu_1).
    \label{eq:second_moment_equation}
\end{equation}
To obtain a closed equation for all the moments, let us
introduce the generating function
\begin{equation}
    \mathcal{M}(t,u)
    =
    \sum_{n=0}^{\infty}
    \mu_n(t)u^n,
    \qquad
    |u|<1.
    \label{eq:moment_generating_function_appendix}
\end{equation}

The two linear contributions to the hierarchy satisfy
\begin{equation}
    \sum_{n=1}^{\infty}
    n\mu_nu^n
    =
    u\partial_u\mathcal{M},
    \label{eq:linear_generating_identity_1}
\end{equation}
and
\begin{equation}
    \sum_{n=1}^{\infty}
    n\mu_{n-1}u^n
    =
    u\partial_u
    \left(
        u\mathcal{M}
    \right).
    \label{eq:linear_generating_identity_2}
\end{equation}

For the nonlinear term, define
\begin{equation}
    C_n
    =
    \sum_{k=0}^{n-2}
    \mu_{n-k-1}
    \left(
        \mu_k-\mu_{k+1}
    \right).
    \label{eq:Cn_definition}
\end{equation}
Since $\mu_0=1$, the summation range is incorporated by writing
\begin{equation}
    \sum_{r=1}^{\infty}
    \mu_ru^r
    =
    \mathcal{M}-1,
\end{equation}
such that
\begin{equation}
    \begin{aligned}
    \sum_{k=0}^{\infty}
    \left(
        \mu_k-\mu_{k+1}
    \right)u^{k+1}
    &=
    u\mathcal{M}
    -
    \left(
        \mathcal{M}-1
    \right)
    =
    1+(u-1)\mathcal{M}.
    \end{aligned}
    \label{eq:difference_moment_series}
\end{equation}
Their product gives
\begin{equation}
    \sum_{n=2}^{\infty}
    C_nu^n
    =
    \left(
        \mathcal{M}-1
    \right)
    \left[
        1+(u-1)\mathcal{M}
    \right].
\label{eq:nonlinear_convolution_generating}
\end{equation}
Applying $u\partial_u$ restores the factor $n$:
\begin{equation}
    \sum_{n=2}^{\infty}
    nC_nu^n
    =
    u\partial_u
    \left\{
        \left(
            \mathcal{M}-1
        \right)
        \left[
            1+(u-1)\mathcal{M}
        \right]
    \right\}.
    \label{eq:nonlinear_n_factor}
\end{equation}
Next, by multiplying the moment hierarchy by $u^n$, summing over $n\geq1$ and after placing the three terms under a common derivative we arrive at
\begin{equation}
    \partial_t\mathcal{M}
    =
    u\partial_u
    \left\{
        \theta(u-1)\mathcal{M}^2
        +
        \left[
            (q-\theta)u+2\theta-1
        \right]\mathcal{M}
    \right\}.  \label{eq:moment_generating_pde_appendix}
\end{equation}

In order to solve this non-linear PDE it is customary to consider the equivalent Cauchy (Green) function of the limiting spectral density 
\begin{equation}
    \mathcal{G}(t,z)
    =
    \int_{0}^{1}
    \frac{\rho_t(\lambda)}{z-\lambda}
    \,\dd\lambda,
    \qquad
    z\in\mathbb{C}\setminus[0,1].
    \label{eq:time_dependent_resolvent}
\end{equation}
For $|z|>1$, a Laurent expansion yields
\begin{equation}
    \mathcal{G}(t,z)
    =
    \frac{1}{z}
    \sum_{n=0}^{\infty}
    \frac{\mu_n(t)}{z^n}.
    \label{eq:resolvent_moment_expansion}
\end{equation}
Hence, both functions are related by
\begin{equation}
    \mathcal{M}(t,u)
    =
    z\mathcal{G}(t,z),
    \qquad
    z=\frac{1}{u}.
    \label{eq:M_to_G_relation}
\end{equation}
Under the transformation $u=1/z$, We therefore obtain
\begin{equation}
    \partial_t\mathcal{G}
    =
    \partial_z
    \left\{
        \left[
            (1-2\theta)z-(q-\theta)
        \right]\mathcal{G}
        +
        \theta z(z-1)\mathcal{G}^2
    \right\}.
\label{eq:nonlinear_resolvent_pde_appendix}
\end{equation}
This is the nonlinear resolvent PDE for the thermodynamic real Jacobi
process. Note that the $\beta$-dependence has disappeared at leading order:
the real and complex processes share the same thermodynamic PDE, while
their finite-size corrections differ.
For the balanced DW geometry,
\begin{equation}
    M=\ell=\frac{L}{2},
\end{equation}
so that
\begin{equation}
    q=\theta=\frac{1}{2}.
\end{equation}
The linear part of the resolvent flux vanishes and
Eq.~\eqref{eq:nonlinear_resolvent_pde_appendix} reduces to
\begin{equation}
    \partial_t\mathcal{G}
    =
    \frac{1}{2}
    \partial_z
    \left[
        z(z-1)\mathcal{G}^2
    \right].
\label{eq:balanced_resolvent_pde_appendix}
\end{equation}
Equivalently, the moment-generating function satisfies
\begin{equation}
    \partial_t\mathcal{M}
    =
    \frac{u}{2}
    \partial_u
    \left[
        (u-1)\mathcal{M}^2
    \right].
\label{eq:balanced_moment_pde_appendix}
\end{equation}
At $t=0$, the DW initial condition is
\begin{equation}
    J(0)=\mathbf{1}_M.
\end{equation}
Therefore,
\begin{equation}
    \mu_n(0)=1
    \qquad
    \text{for every }n\geq0,
\end{equation}
and
\begin{equation}
    \mathcal{M}(0,u)
    =
    \frac{1}{1-u},
    \qquad
    \mathcal{G}(0,z)
    =
    \frac{1}{z-1}.   
\label{eq:balanced_initial_generating_resolvent}
\end{equation}

\section{Solution by the method of characteristics.}
\label{app:characteristics}
In this appendix we solve Eq.~\eqref{eq:balanced_resolvent_pde_appendix} directly by
the method of characteristics. Characteristic methods for the free-Jacobi process equations have appeared previously in several closely related
settings~\cite{Demni2008,DemniHamdiHmidi2012,DemniHamdi2025Moments}.
In particular, characteristic curves have been obtained for the balanced
case $q=\theta=1/2$ and, more recently, for free Jacobi processes associated with projections
of equal rank~\cite{demni2025spectral}. The latter formulation encodes
the flow through the inverse of the transform associated with free unitary Brownian motion. A complementary formulation in
terms of the free $S$ transform and its characteristic equations was developed
in Ref.~\cite{demni2025finite}.

Here we give a self-contained derivation in the resolvent variables used in the
present work. Besides recovering the known balanced structure, this form will
be useful below because it admits a direct extension to unequal subsystem and
projector fractions.

Starting from Eq.\eqref{eq:balanced_resolvent_pde_appendix}, we can expand the derivative and obtain
\begin{equation}
    \partial_t\mathcal{G}
    +
    z(1-z)\mathcal{G}\,\partial_z\mathcal{G}
    =
    \left(z-\frac{1}{2}\right)\mathcal{G}^2.
    \label{eq:balanced_quasilinear_G}
\end{equation}
Therefore, the characteristic equations are 
\begin{equation}
    \dot z
    =
    z(1-z)\mathcal{G},
    \qquad
    \dot{\mathcal{G}}
    =
    \left(z-\frac{1}{2}\right)\mathcal{G}^2.
    \label{eq:balanced_characteristic_equations}
\end{equation}
We label each characteristic by its initial position $\zeta$,
\begin{equation}
    z(0)=\zeta,
    \qquad
    \mathcal{G}(0)
    =
    \frac{1}{\zeta-1}.
    \label{eq:characteristic_initial_data}
\end{equation}
Surprisingly the coupled equations possess a simple first integral
\begin{align}
    \frac{\dd}{\dd t}
    \left[
        z(1-z)\mathcal{G}^2
    \right]
    &=
    (1-2z)\dot z\,\mathcal{G}^2
    +
    2z(1-z)\mathcal{G}\dot{\mathcal{G}}
    =0.
\end{align}
Using the initial condition~\eqref{eq:characteristic_initial_data} the conserved quantity takes the value
\begin{equation}
    z(1-z)\mathcal{G}^2
    =
    \frac{\zeta}{1-\zeta}.
\label{eq:characteristic_first_integral}
\end{equation}
The remaining task is to solve for $z$ in terms of the characteristic $\zeta$. Combining Eq.~\eqref{eq:characteristic_first_integral} with the first
equation in Eq.~\eqref{eq:balanced_characteristic_equations} gives
\begin{equation}
    \dot z^{\,2}
    =
    \frac{\zeta}{1-\zeta}
    z(1-z).
    \label{eq:z_characteristic_squared}
\end{equation}
The initial velocity is
\begin{equation}
    \dot z(0)
    =
    \zeta(1-\zeta)
    \frac{1}{\zeta-1}
    =
    -\zeta,
\end{equation}
which selects the branch
\begin{equation}
    \dot z
    =
    -
    \sqrt{
        \frac{\zeta}{1-\zeta}
        z(1-z)
    }.
    \label{eq:z_characteristic_branch}
\end{equation}

It is convenient to parametrize the initial and final spectral
coordinates by
\begin{equation}
    \zeta=\cos^2\eta,
    \qquad
    z=\cos^2 x.
    \label{eq:angular_characteristic_variables}
\end{equation}
On the real branch $0<\eta<\pi/2$,
\begin{equation}
    \sqrt{\frac{\zeta}{1-\zeta}}
    =
    \cot\eta.
\end{equation}
Substituting Eq.~\eqref{eq:angular_characteristic_variables} into
Eq.~\eqref{eq:z_characteristic_branch} yields the constant-velocity
equation
\begin{equation}
    \dot x=\frac{1}{2}\cot\eta.
\end{equation}
Since $x(0)=\eta$, the characteristic map is
\begin{equation}
    x(t,\eta)
    =
    \eta+\frac{t}{2}\cot\eta,    
\label{eq:angular_characteristic_map}
\end{equation}
and hence
\begin{equation}
    z(t,\eta)
    =
    \cos^2
    \left(
        \eta+\frac{t}{2}\cot\eta
    \right).
\label{eq:z_parametric_characteristic}
\end{equation}

The resolvent along the same characteristic follows directly from
$\mathcal{G}=\dot z/[z(1-z)]$. Using
$z=\cos^2 x$, one finds
\begin{equation}
    \mathcal{G}(t,\eta)
    =
    -
    \frac{2\cot\eta}
    {\sin\!\left(2\eta+t\cot\eta\right)}
    =
    -
    \frac{2\cot\eta}{\sin 2x}.
\label{eq:G_parametric_characteristic}
\end{equation}
Equations~\eqref{eq:z_parametric_characteristic} and
\eqref{eq:G_parametric_characteristic} give a parametric solution of the
balanced resolvent PDE.

The above expression is already an exact solution to the resolvent PDE, but in order to obtain a representation suitable for series expansion, we introduce the variables
\begin{equation}
    w=e^{i\eta},
    \qquad
    v=e^{ix}.
\end{equation}
Using
\begin{equation}
    \cot\eta
    =
    i\frac{w^2+1}{w^2-1},
\end{equation}
Eq.~\eqref{eq:angular_characteristic_map} becomes
\begin{equation}
    v
    =
    w
    \exp
    \left[
        -\frac{t}{2}
        \frac{w^2+1}{w^2-1}
    \right].
    \label{eq:wv_characteristic}
\end{equation}
To remove the redundant square, we further define
\begin{equation}
    y=w^2,
    \qquad
    s=v^2.
\end{equation}
Then following expression appears
\begin{equation}
    s
    =
    y
    \exp
    \left[
        t\frac{1+y}{1-y}
    \right],
    \label{eq:sy_characteristic}
\end{equation}
or equivalently
\begin{equation}
    y
    =
    s
    \exp
    \left[
        -t\frac{1+y}{1-y}
    \right].
    \label{eq:y_implicit_equation}
\end{equation}

Equation~\eqref{eq:G_parametric_characteristic} becomes
\begin{equation}
    \mathcal{G}(t,s)
    =
    -
    \frac{4s}{s^2-1}
    \frac{1+y(s)}{1-y(s)}.
    \label{eq:G_sy_form}
\end{equation}
Returning to the moment-generating variable,
let $u=1/z$ be the moment-generating variable introduced in
Eq.~\eqref{eq:moment_generating_function_appendix}. Since
$z=\cos^2x$ and $s=e^{2ix}$,
\begin{equation}
    z
    =
    \frac{(1+s)^2}{4s},
\end{equation}
and therefore
\begin{equation}
    u
    =
    \frac{4s}{(1+s)^2}.
    \label{eq:u_s_conformal_map}
\end{equation}
The branch analytic at $u=0$ is
\begin{equation}
    s(u)
    =
    \frac{1-\sqrt{1-u}}
         {1+\sqrt{1-u}}
    =
    \frac{u}
    {\left(1+\sqrt{1-u}\right)^2},
    \label{eq:s_of_u}
\end{equation}
for which $s(u)\sim u/4$ as $u\rightarrow0$. The same map gives the useful identity
\begin{equation}
    \frac{1+s}{1-s}
    =
    \frac{1}{\sqrt{1-u}}.
    \label{eq:s_root_identity}
\end{equation}

Define
\begin{equation}
    H(y)=\frac{1+y}{1-y},
\end{equation}
and using $\mathcal{M}=z\mathcal{G}$ together with
Eqs.~\eqref{eq:G_sy_form} and \eqref{eq:s_root_identity}, one obtains
\begin{equation}
    \mathcal{M}(t,u)
    =
    \frac{1}{\sqrt{1-u}}
    H\!\left(y(s(u))\right).
    \label{eq:M_H_representation}
\end{equation}

Equation~\eqref{eq:y_implicit_equation} has the canonical implicit form
\begin{equation}
    y=s\phi(y),
    \qquad
    \phi(y)
    =
    \exp[-tH(y)].
    \label{eq:lagrange_canonical_form}
\end{equation}
For an analytic function $F$, the Lagrange--B\"urmann inversion formula~\cite{Comtet1974}
states that
\begin{equation}
    [s^k]F(y(s))
    =
    \frac{1}{k}
    [\xi^{k-1}]
    F'(\xi)\phi(\xi)^k,
    \qquad k\geq1,
    \label{eq:lagrange_burmann_formula}
\end{equation}
where $[s^k]$ denotes the $k$th order coefficient of the Laurent expansion for $F(s)$.
Applying Eq.~\eqref{eq:lagrange_burmann_formula} to $F=H$, we have
\begin{equation}
    H'(\xi)=\frac{2}{(1-\xi)^2},
\end{equation}
and
\begin{align}
    \phi(\xi)^k
    &=
    e^{-kt}
    \exp
    \left[
        -\frac{2kt\,\xi}{1-\xi}
    \right].
    \label{eq:phi_k_laguerre_form}
\end{align}
We now use the generating function of the associated Laguerre
polynomials,
\begin{equation}
    \frac{1}{(1-\xi)^{\alpha+1}}
    \exp
    \left[
        -\frac{x\xi}{1-\xi}
    \right]
    =
    \sum_{m=0}^{\infty}
    L_m^{(\alpha)}(x)\xi^m.
    \label{eq:laguerre_generating_function}
\end{equation}
For $\alpha=1$ and $x=2kt$,
\begin{equation}
    [\xi^{k-1}]
    H'(\xi)\phi(\xi)^k
    =
    2e^{-kt}
    L_{k-1}^{(1)}(2kt).
\end{equation}
Consequently,
\begin{equation}
    H(y(s))
    =
    1
    +
    2
    \sum_{k=1}^{\infty}
    \frac{1}{k}
    L_{k-1}^{(1)}(2kt)
    e^{-kt}
    s^k.
    \label{eq:H_laguerre_series}
\end{equation}
Substituting Eq.~\eqref{eq:H_laguerre_series} into
Eq.~\eqref{eq:M_H_representation} yields
\begin{equation}
    \mathcal{M}(t,u)
    =
    \frac{1}{\sqrt{1-u}}
    \left[
        1
        +
        2
        \sum_{k=1}^{\infty}
        \frac{1}{k}
        L_{k-1}^{(1)}(2kt)
        e^{-kt}
        s(u)^k
    \right],
    \label{eq:M_exact_laguerre_representation}
\end{equation}
where $s(u)$ is given by Eq.~\eqref{eq:s_of_u}.

Finally, for the extraction of the moments, the time-independent contribution follows from
\begin{equation}
    \frac{1}{\sqrt{1-u}}
    =
    \sum_{n=0}^{\infty}
    \frac{1}{4^n}
    \binom{2n}{n}
    u^n,
    \label{eq:central_binomial_expansion}
\end{equation}
so that
\begin{equation}
    \mu_n^{(\mathrm{stat})}
    =
    \frac{1}{4^n}
    \binom{2n}{n}.
\end{equation}

For the dynamical part, define
\begin{equation}
    C_{n,k}
    =
    [u^n]
    \frac{s(u)^k}{\sqrt{1-u}}.
    \label{eq:Cnk_definition}
\end{equation}
Cauchy's coefficient formula gives
\begin{equation}
    C_{n,k}
    =
    \frac{1}{2\pi i}
    \oint
    \frac{\dd u}{u^{n+1}}
    \frac{s(u)^k}{\sqrt{1-u}},
    \label{eq:Cnk_cauchy}
\end{equation}
where the contour is a sufficiently small positively oriented circle
around $u=0$.

Using
\begin{equation}
    u=\frac{4s}{(1+s)^2},
    \qquad
    \frac{\dd u}{\dd s}
    =
    4\frac{1-s}{(1+s)^3},
    \qquad
    \frac{1}{\sqrt{1-u}}
    =
    \frac{1+s}{1-s},
\end{equation}
Eq.~\eqref{eq:Cnk_cauchy} reduces to
\begin{equation}
    C_{n,k}
    =
    \frac{1}{4^n}
    \frac{1}{2\pi i}
    \oint
    s^{k-n-1}
    (1+s)^{2n}
    \dd s.
    \label{eq:Cnk_s_contour}
\end{equation}
Expanding
\begin{equation}
    (1+s)^{2n}
    =
    \sum_{m=0}^{2n}
    \binom{2n}{m}s^m,
\end{equation}
the residue at $s=0$ is selected by
\begin{equation}
    m+k-n-1=-1,
\end{equation}
or
\begin{equation}
    m=n-k.
\end{equation}
Since $m\geq0$, the contribution vanishes for $k>n$. Therefore,
\begin{equation}
    C_{n,k}
    =
    \frac{1}{4^n}
    \binom{2n}{n-k},
    \qquad
    1\leq k\leq n.
    \label{eq:Cnk_final}
\end{equation}

Combining Eqs.~\eqref{eq:M_exact_laguerre_representation},
\eqref{eq:central_binomial_expansion}, and
\eqref{eq:Cnk_final}, we obtain
\begin{equation}
    \mu_n(t)
    =
    \frac{1}{4^n}
    \binom{2n}{n}
    +
    \frac{2}{4^n}
    \sum_{k=1}^{n}
    \binom{2n}{n-k}
    \frac{1}{k}
    L_{k-1}^{(1)}(2kt)
    e^{-kt}.
\label{eq:balanced_exact_moments_appendix}
\end{equation}
This reproduces the exact balanced free-Jacobi moment formula of
Refs.~\cite{Demni2008,DemniHamdiHmidi2012} and the form used in
Ref.~\cite{BernardPiroliScopa2026}.  Lastly, we note that recent works~\cite{DemniHamdi2025Moments} have successfully obtained finite-time moment expressions for arbitrary free-Jacobi parameters, together with finite recurrence relations for their coefficients.

Several checks follow immediately. Taking $n=1$,
\begin{equation}
    \mu_1(t)
    =
    \frac{1}{2}
    \left(
        1+e^{-t}
    \right),
\end{equation}
which solves
\begin{equation}
    \partial_t\mu_1
    =
    -\mu_1+\frac{1}{2},
    \qquad
    \mu_1(0)=1.
\end{equation}
In the stationary limit,
\begin{equation}
    \lim_{t\rightarrow\infty}\mu_n(t)
    =
    \frac{1}{4^n}
    \binom{2n}{n},
\end{equation}
which are precisely the moments of the arcsine distribution. Finally,
at $t=0$, Eq.~\eqref{eq:y_implicit_equation} gives $y=s$, so that
\begin{equation}
    H(y)=H(s)=\frac{1}{\sqrt{1-u}},
\end{equation}
and Eq.~\eqref{eq:M_H_representation} reduces to
\begin{equation}
    \mathcal{M}(0,u)=\frac{1}{1-u},
\end{equation}
recovering $\mu_n(0)=1$ for every $n$.

% ============================================================
% Appendix: arbitrary q and theta
% ============================================================

\section{Characteristic solution and OSEE at arbitrary $q$ and $\theta$}
\label{app:general_characteristics}

In this appendix we extend the characteristic construction to arbitrary subsystem fraction
\begin{equation}
    q=\frac{\ell}{L}
\end{equation}
and filling fraction
\begin{equation}
    \theta=\frac{M}{L}.
\end{equation}

Finite-time moments of the free Jacobi process for arbitrary parameters have
recently been obtained by Demni and Hamdi, with the
general coefficients organized through finite recurrence
relations~\cite{DemniHamdi2025Moments}.  Complementary characteristic
descriptions are known in special parameter regimes.  In particular,
Ref.~\cite{demni2025spectral}  determines the characteristic curves for
projections of equal rank, while
Ref.~\cite{demni2025finite} derives a PDE for the free
$S$ transform at general Jacobi parameters and analyzes its characteristic
flow explicitly in a special parameter regime.

Our approach here is complementary.  Rather than reconstructing the
finite-time measure from its moments, we solve the nonlinear resolvent
equation directly. We remark that our solution yields a parametric representation valid for
arbitrary $(q,\theta)$, including different-rank configurations, and allows the
OSEE to be evaluated directly from the resolvent without explicitly summing
the full finite-time moment hierarchy.

% ------------------------------------------------------------
% ------------------------------------------------------------

At $t=0$ the orthogonal rotation is the identity, $R(0)=I_L$. The matrix
$X$ is obtained by restricting the first $M$ columns of $R$ to the first
$\ell$ rows. Therefore the initial Jacobi matrix
\begin{equation}
    J(0)=X(0)^T X(0)
\end{equation}
has only eigenvalues $0$ and $1$.
Introducing
\begin{equation}
    r=\min\left(1,\frac{q}{\theta}\right),
\end{equation}
the normalized initial spectral measure is
\begin{equation}
    \rho_0(\lambda)
    =
    (1-r)\delta(\lambda)
    +
    r\,\delta(\lambda-1).
    \label{eq:rho0_general}
\end{equation}
Consequently,
\begin{equation}
    \cG_0(z)
    =
    \frac{1-r}{z}
    +
    \frac{r}{z-1}
    =
    \frac{z-1+r}{z(z-1)}.
    \label{eq:G0_general}
\end{equation}

For $q\geq\theta$, one has $r=1$ and hence
\begin{equation}
    \cG_0(z)=\frac{1}{z-1}.
\end{equation}

% ------------------------------------------------------------
% ------------------------------------------------------------
Expanding the derivative in the r.h.s of  Eq.~\eqref{eq:resolvent_pde} gives
\begin{equation}
    \partial_t \cG
    +
    \left[
        (2\theta-1)z+(q-\theta)
        +2\theta z(1-z)\cG
    \right]\partial_z \cG
    =
    (1-2\theta)\cG+\theta(2z-1)\cG^2.
\end{equation}
The characteristic equations are therefore
\begin{align}
    \dot z
    &=
    (2\theta-1)z+(q-\theta)
    +2\theta z(1-z)\cG,
    \label{eq:char_z_general}
    \\
    \dot \cG
    &=
    (1-2\theta)\cG
    +\theta(2z-1)\cG^2.
    \label{eq:char_G_general}
\end{align}
Since the vector field is divergence free,
\begin{equation}
    \partial_z\dot z+\partial_\cG\dot \cG=0,
\end{equation}
we can rewrite the above equations locally in Hamiltonian form,
\begin{equation}
    \dot z=\partial_\cG\mathcal H,
    \qquad
    \dot \cG=-\partial_z\mathcal H,
\end{equation}
with
\begin{equation}
    \mathcal H(z,\cG)
    =
    \left[(2\theta-1)z+(q-\theta)\right]\cG
    +
    \theta z(1-z)\cG^2.
    \label{eq:H_general}
\end{equation}
Hence $\mathcal H$ is conserved along each characteristic. Choosing the initial characteristic coordinate be
\begin{equation}
    z(0)=\zeta,
    \qquad
    \cG(0)=\cG_0(\zeta),
\end{equation}
the corresponding conserved energy is given by
\begin{equation}
    \mathcal E(\zeta)
    =
    \frac{
        (\zeta-1+r)
        \left[(1-\theta)\zeta-q+\theta r\right]
    }{
        \zeta(1-\zeta)
    }.
    \label{eq:E_general}
\end{equation}

Introduce
\begin{equation}
    B(z)=(2\theta-1)z+(q-\theta),
    \qquad
    C(z)=\theta z(1-z),
\end{equation}
such that
\begin{equation}
    \mathcal E=B(z)\cG+C(z)\cG^2,
\end{equation}
and
\begin{equation}
    \dot z=B(z)+2C(z)\cG.
\end{equation}
Completing the square gives
\begin{equation}
    \dot z^{\,2}
    =
    B(z)^2+4C(z)\mathcal E.
    \label{eq:zdot_squared_general}
\end{equation}
Note that the right-hand side is only quadratic in $z$:
\begin{equation}
    \dot z^{\,2}
    =
    \mathsf A(\mathcal E)z^2
    +
    2\mathsf B(\mathcal E)z
    +
    (q-\theta)^2,
    \label{eq:quadratic_z}
\end{equation}
where
\begin{align}
    \mathsf A(\mathcal E)
    &=
    (2\theta-1)^2-4\theta\mathcal E,
    \\
    \mathsf B(\mathcal E)
    &=
    (2\theta-1)(q-\theta)+2\theta\mathcal E.
\end{align}
Differentiating Eq.~\eqref{eq:quadratic_z} with respect to time yields
\begin{equation}
    \ddot z
    =
    \mathsf A(\mathcal E)z+\mathsf B(\mathcal E).
    \label{eq:z_second_order}
\end{equation}
Thus the apparently nonlinear first-order problem is equivalent to a linear
second-order equation along every characteristic.

Introducing
\begin{equation}
    s^2=\mathsf A(\mathcal E),
    \label{eq:s_definition}
\end{equation}
we choose the branch such that $s\rightarrow1$ for $\zeta\rightarrow\infty$.
Furthermore introduce
\begin{equation}
    q_<\equiv\min(q,\theta),
    \qquad
    q_>\equiv\max(q,\theta),
\end{equation}
and
\begin{equation}
    \kappa=q_<\left(1-q_>\right),
    \qquad
    \delta=q_>-q_< = |q-\theta|.
\end{equation}
Using Eq.~\eqref{eq:E_general}, one finds the important simplification
\begin{equation}
    s^2
    =
    1+\frac{4\kappa}{\zeta-1}.
    \label{eq:s_zeta_relation}
\end{equation}
Equivalently,
\begin{equation}
    \zeta
    =
    1+\frac{4\kappa}{s^2-1}.
    \label{eq:zeta_s_relation}
\end{equation}

The center of the linear equation is
\begin{equation}
    z_c(s)
    =
    \frac12
    -
    \frac{(2q-1)(2\theta-1)}{2s^2}.
    \label{eq:zc_general}
\end{equation}
Moreover, the initial velocity simplifies to
\begin{equation}
    \dot z(0)=\delta-\zeta.
\label{eq:initial_velocity_general}
\end{equation}
The characteristic solution can therefore be written as
\begin{equation}
    z(t,s)
    =
    z_c(s)+D(s)e^{st}+F(s)e^{-st},
    \label{eq:z_DF_general}
\end{equation}
where
\begin{align}
    D(s)
    &=
    \frac{
        (s-1)(s+2q_>-1)(s+1-2q_<)
    }{
        4s^2(s+1)
    },
    \label{eq:D_general}
    \\
    F(s)
    &=
    \frac{
        (s+1)(s-2q_>+1)(s-1+2q_<)
    }{
        4s^2(s-1)
    }.
    \label{eq:F_general}
\end{align}

The factors $(s\mp1)$ appearing in Eqs.~\eqref{eq:D_general} and
\eqref{eq:F_general} suggest defining the expression
\begin{equation}
    y
    =
    \frac{s-1}{s+1}e^{st}.
    \label{eq:y_general}
\end{equation}
Then Eq.~\eqref{eq:z_DF_general} collapses to the elegant form
\begin{equation}
    z(t,s)
    =
    \frac{1+y}{4s^2y}
    \left[
        P_+(s)y+P_-(s)
    \right],
    \label{eq:z_sy_general}
\end{equation}
with
\begin{align}
    P_+(s)
    &=
    (s+2q_>-1)(s+1-2q_<),
    \\
    P_-(s)
    &=
    (s-2q_>+1)(s-1+2q_<).
\end{align}

From Eq.~\eqref{eq:char_z_general}, the resolvent along a characteristic is
\begin{equation}
    \cG
    =
    \frac{
        \dot z-(2\theta-1)z-(q-\theta)
    }{
        2\theta z(1-z)
    }.
    \label{eq:G_from_zdot}
\end{equation}

For definiteness, consider first $q\geq\theta$. Define
\begin{equation}
    a=1-2\theta,
    \qquad
    b=2q-1.
    \label{eq:ab_definition}
\end{equation}
Substitution of Eq.~\eqref{eq:z_sy_general} into
Eq.~\eqref{eq:G_from_zdot} gives
\begin{equation}
    \cG(t,s)
    =
    -
    \frac{
        2s^2y
    }{
        \theta(1+y)
        \left[
            s(y-1)+b(1+y)
        \right]
    }.
    \label{eq:G_sy_general}
\end{equation}

A further simplification is obtained by defining
\begin{equation}
    h
    =
    s\frac{1-y}{1+y}.
    \label{eq:h_definition_general}
\end{equation}
Since
\begin{equation}
    y=\frac{s-h}{s+h},
\end{equation}
Eq.~\eqref{eq:y_general} is equivalent to
\begin{equation}
    \frac{s-h}{s+h}
    =
    \frac{s-1}{s+1}e^{st}.
    \label{eq:implicit_sh_general}
\end{equation}
Solving explicitly for $h$ gives
\begin{equation}
    h(s,t)
    =
    s\,
    \frac{
        1+s+(1-s)e^{st}
    }{
        1+s-(1-s)e^{st}
    }.
    \label{eq:h_explicit_general}
\end{equation}

In terms of $(s,h)$, both the spectral coordinate and the resolvent become
particularly simple:
\begin{equation}
    z
    =
    1-\frac{(a-h)(b-h)}{h^2-s^2},
    \label{eq:z_sh_general}
\end{equation}
and
\begin{equation}
    (z-1)\cG
    =
    \frac{h-a}{2\theta}.
    \label{eq:G_sh_general}
\end{equation}
Equations~\eqref{eq:implicit_sh_general},
\eqref{eq:z_sh_general}, and \eqref{eq:G_sh_general} constitute an exact
parametric solution of the thermodynamic resolvent equation for
$q\geq\theta$.
Equivalently, introducing $u=1/z$ and the moment-generating function
$\mathcal M=z\cG$, one has
\begin{align}
    u
    &=
    \frac{
        h^2-s^2
    }{
        (a+b)h-ab-s^2
    },
    \label{eq:u_sh_general}
    \\
    \mathcal M(t,u)
    &=
    \frac{
        h-a
    }{
        2\theta(1-u)
    }.
    \label{eq:M_sh_general}
\end{align}
For $q<\theta$, the matrix $X^TX$ contains an additional set of exact zero
modes, while its nonzero eigenvalues coincide with those of $XX^T$. Hence
\begin{equation}
    \cG_{q,\theta}(z)
    =
    \left(1-\frac{q}{\theta}\right)\frac{1}{z}
    +
    \frac{q}{\theta}\,
    \cG_{\theta,q}(z).
    \label{eq:resolvent_duality}
\end{equation}
Here, the first term accounts for the $\lambda=0 $ eigenvalues, and for the second term we used that, at the distribution level, the Brownian motion generator satisfies $R_t= R_t^T$ since the tangential generators $\Delta K_t$ are independent and identically distributed gaussian variables.
Since the binary entropy vanishes at $\lambda=0$, the total OSEE is symmetric
under exchanging the two fractions,
\begin{equation}
    S_{\rm op}(q,\theta;t)
    =
    S_{\rm op}(\theta,q;t).    \label{eq:entropy_qtheta_symmetry}
\end{equation}
Although we obtained an exact solution for the normalized resolvent $\cG$ a simple formula for the moments does not follow immediately. Instead, since we are concerned with computing the thermodynamic limit dynamics of OSEE, a particularly useful observation is the integral representation
\begin{equation}
    -\log x
    =
    \int_0^\infty
    d\tau
    \left(
        \frac{1}{\tau+x}
        -
        \frac{1}{\tau+1}
    \right),
    \label{eq:log_integral_rep}
\end{equation}
the binary entropy can be rewritten as

\begin{equation}
    \rm h_2(\lambda)
    =
    \int_0^\infty d\tau
    \left[
        \frac{\lambda}{\tau+\lambda}
        +
        \frac{1-\lambda}{\tau+1-\lambda}
        -
        \frac{1}{\tau+1}
    \right].
\label{eq:binary_entropy_integral}
\end{equation}
Moreover, by writing the resolvent as the Stieltjes transform of the Jacobi spectral
measure
\begin{equation}
    \cG(t,z)
    =
    \frac{1}{M}
    \Tr\frac{1}{z-J(t)}
    =
    \int_0^1
    \frac{\rho_t(\lambda)}{z-\lambda}\,d\lambda,
    \label{eq:stieltjes_general}
\end{equation}
and using
\begin{align}
    \int_0^1d\lambda\,
    \rho_t(\lambda)
    \frac{\lambda}{\tau+\lambda}
    &=
    1+\tau \cG(t,-\tau),
    \label{eq:first_resolvent_identity}
    \\
    \int_0^1d\lambda\,
    \rho_t(\lambda)
    \frac{1-\lambda}{\tau+1-\lambda}
    &=
    1-\tau \cG(t,1+\tau),
    \label{eq:second_resolvent_identity}
\end{align}
the thermodynamic OSEE becomes
\begin{equation}
    \frac{S_{\rm op}(t)}{L}
    =
    \theta
    \int_0^\infty d\tau\,
    \left[
        2-\frac{1}{1+\tau}
        +
        \tau\left(
            \cG(t,-\tau)-\cG(t,1+\tau)
        \right)
    \right].
\label{eq:entropy_final_resolvent}
\end{equation}
This expression involves the resolvent only outside its physical spectral
support, namely on $z<0$ and $z>1$. Hence, no reconstruction of the
time-dependent spectral density or analytic continuation onto the spectral
cut is required.

For $q\geq\theta$, let $s_-(\tau)$ and $s_+(\tau)$ denote the two physical
real branches determined by
\begin{equation}
    z(s_-(\tau),t)=-\tau,
    \qquad
    z(s_+(\tau),t)=1+\tau,
    \label{eq:exterior_branches}
\end{equation}
which are continuously connected to $s=1^-$ and $s=1^+$, respectively, as
$\tau\rightarrow\infty$.
Using Eq.~\eqref{eq:G_sh_general}, we obtain
\begin{align}
    \tau G(t,1+\tau)
    &=
    \frac{
        h_+(\tau)-a
    }{
        2\theta
    },
    \\
    \tau G(t,-\tau)
    &=
    -
    \frac{\tau}{1+\tau}
    \frac{
        h_-(\tau)-a
    }{
        2\theta
    },
\end{align}
where we introduced the compact notation 
\begin{equation}
    h_\pm(\tau,t)
    \equiv
    h(s_\pm(\tau,t),t).
\end{equation}
Therefore, the final entropy formula can be written entirely in terms of the
two real characteristic branches:
\begin{equation}
    \frac{S_{\rm op}(t)}{L}
    =
    \int_0^\infty d\tau\,
    \left\{
        \theta
        \left(
            2-\frac{1}{1+\tau}
        \right)
        -
        \frac{1}{2}
        \left[
            \frac{\tau}{1+\tau}
            \left(h_--a\right)
            +
            \left(h_+-a\right)
        \right]
    \right\}.
    \label{eq:entropy_final_h}
\end{equation}
Equation~\eqref{eq:entropy_final_h} is an exact one-dimensional quadrature
for the thermodynamic OSEE at arbitrary subsystem and filling fractions due to the invariance of OSEE under $q \leftrightarrow \theta$.
Its numerical evaluation can be performed efficiently without solving the
implicit characteristic equations independently at every value of $\tau$. Since the two physical branches are defined implicitly by
\begin{equation}
    z(s_+(\tau),t)=1+\tau,
    \qquad
    z(s_-(\tau),t)=-\tau,
    \label{eq:entropy_branch_definition}
\end{equation}
differentiating the above relations at fixed $t$ gives
\begin{equation}
    \frac{d s_+}{d\tau}
    =
    \frac{1}
    {\partial_s z(s_+(\tau),t)},
    \qquad
    \frac{d s_-}{d\tau}
    =
    -
    \frac{1}
    {\partial_s z(s_-(\tau),t)}.
    \label{eq:entropy_branch_ode_tau}
\end{equation}
Thus, after fixing the physical roots at a single sufficiently large value
$\tau_{\max}$, the complete branches can be obtained by numerical
continuation rather than by repeated root searches.  The initial roots are
chosen as the solutions continuously connected to
\begin{equation}
    s_+\rightarrow1^+,
    \qquad
    s_-\rightarrow1^-,
    \qquad
    \tau\rightarrow\infty.
\end{equation}
In practice it is advantageous to introduce the logarithmic integration
coordinate
\begin{equation}
    x=\log\tau,
    \qquad
    \tau=e^x,
\end{equation}
to control the large values of
$\tau$ in the entropy integral.  Eq.~\eqref{eq:entropy_branch_ode_tau} then becomes
\begin{equation}
    \frac{d s_+}{dx}
    =
    \frac{e^x}
    {\partial_s z(s_+(x),t)},
    \qquad
    \frac{d s_-}{dx}
    =
    -
    \frac{e^x}
    {\partial_s z(s_-(x),t)}.
    \label{eq:entropy_branch_ode_log}
\end{equation}
Writing the integrand of Eq.~\eqref{eq:entropy_final_h} as
$\mathcal I(\tau,t)$, the finite numerical contribution is
\begin{equation}
    \frac{S_{\rm op}^{\rm mid}(t)}{L}
    =
    \int_{\log\tau_{\min}}^{\log\tau_{\max}}
    dx\,
    e^x\,
    \mathcal I(e^x,t).
    \label{eq:entropy_log_quadrature}
\end{equation}
The remaining intervals can be controlled analytically.  Since the
integrand is regular at the lower endpoint, the contribution from
$0<\tau<\tau_{\min}$ vanishes with $\tau_{\min}$.  At large $\tau$ one finds
\begin{equation}
    \mathcal I(\tau,t)
    =
    \frac{C(t)}{\tau^2}
    +
    O(\tau^{-3}),
\end{equation}
so that the omitted upper tail can be estimated as
\begin{equation}
    \int_{\tau_{\max}}^\infty
    d\tau\,\mathcal I(\tau,t)
    =
    \tau_{\max}\,
    \mathcal I(\tau_{\max},t)
    +
    O(\tau_{\max}^{-2}).
\end{equation}

\nocite{*}
\bibliographystyle{iopart-num}
\bibliography{references}

\end{document}